\documentclass[aps,a4paper,twocolumn,nofootinbib,eqsecnum]{revtex4}  %
   
\usepackage{graphicx} 
\usepackage{mathrsfs,amsmath,amsfonts,amssymb,multirow}

\newcommand{\beq}{\begin{equation}}
\newcommand{\eeq}{\end{equation}}

\begin{document}

\title{Scattering of tidally interacting bodies in post-Minkowskian gravity}

\author{Donato Bini$^{1,2}$, Thibault Damour$^3$, Andrea Geralico$^1$}
  \affiliation{
$^1$Istituto per le Applicazioni del Calcolo ``M. Picone,'' CNR, I-00185 Rome, Italy\\
$^2$INFN, Sezione di Roma Tre, I-00146 Rome, Italy\\
$^3$Institut des Hautes \'Etudes Scientifiques, 91440 Bures-sur-Yvette, France.
}

\date{\today}

\begin{abstract}
The post-Minkowskian approach to gravitationally interacting binary systems
({\it i.e.}, perturbation theory in $G$, without assuming small velocities) is extended to the computation
of the dynamical effects induced by the tidal deformations of two extended bodies, such as neutron stars.
Our derivation applies general properties of perturbed actions to the effective field theory description of 
tidally interacting bodies. We compute several tidal invariants (notably the integrated quadrupolar
and octupolar actions) at the first post-Minkowskian order. The corresponding contributions to the scattering
angle are derived.
\end{abstract}

\maketitle

\section{Introduction}

The post-Minkowskian (PM) approach to gravitational interaction, which was pioneered some time ago \cite{Bertotti56,bert-pleb,Portilla:1980uz,Westpfahl:1979gu,Bel:1981be,Westpfahl:1985,Westpfahl:1987,Ledvinka:2008tk}, 
has been recently revived \cite{Damour:2016gwp,Damour2018} and has undergone many developments both in classical gravity 
\cite{Vines:2017hyw,Bini:2017xzy,Bini:2018ywr,Vines:2018gqi,Antonelli:2019ytb}, and in the connection between classical gravity and quantum scattering amplitudes \cite{Cachazo:2017jef,Guevara:2017csg,Bjerrum-Bohr:2018xdl,Cheung:2018wkq,KoemansCollado:2019ggb,Bern:2019nnu,Bern:2019crd,Guevara:2018wpp,Chung:2018kqs,Guevara:2019fsj,Siemonsen:2019dsu,Kalin:2019rwq,Bjerrum-Bohr:2019kec,DiVecchia:2019myk,DiVecchia:2019kta,Kalin:2019inp,Blumlein:2019bqq,Bjerrum-Bohr:2019nws,Maybee:2019jus,Cristofoli:2019ewu,Ciafaloni:2018uwe}. The aim of  the present paper is to extend the post-Minkowskian approach to tidal effects in binary systems.

Tidal interactions are expected to play an important role in driving the dynamics of the last orbits of coalescing binary systems comprising at least one neutron star.
Up to now, tidal effects in binary systems have been studied within either: i) the  post-Newtonian (PN) approach \cite{Damour:1992qi,Damour:1993zn,Vines:2010ca}; ii)  numerical relativity (see e.g., \cite{Shibata:1999hn,Baiotti:2008ra,Bernuzzi:2012ci,Bernuzzi:2014owa}); and iii) the gravitational self-force (SF) approach 
\cite{Bini:2014zxa,Kavanagh:2015lva,Nolan:2015vpa}. In addition, it was found useful \cite{Damour:2009wj} to 
transcribe the results  of the latter approaches within the effective one-body (EOB) formalism \cite{Buonanno:1998gg,Buonanno:2000ef,Damour:2000we}.
 
The state-of-the-art of our analytical knowledge of tidal interactions in non-spinning comparable-mass binary systems is presently limited to the second PN approximation \cite{Bini:2012gu}, while,  in the limiting situation of  extreme-mass-ratio systems, SF theory has obtained high-order PN results in the framework of linear perturbation theory \cite{Bini:2014zxa,Bini:2015bla,Bini:2015mza}.

The starting point of our present computation is 
the  effective field theory description of the dynamics of gravitationally interacting extended bodies \cite{Damour:1995kt,Damour:1998jk,Goldberger:2004jt,Damour:2009wj}. 
This approach will be briefly recalled in Sec. III.
It describes finite size effects  by adding to the point-mass action of a two-body system certain non-minimal worldline couplings,
defined as integrals of tidal invariants along the worldlines of the bodies. 
The coefficients appearing in front of these non-minimal worldline couplings are certain tidal polarizability parameters (linked to \lq\lq Love numbers"), which can be computed  given some equation of state for the nuclear matter \cite{Damour:1982wm,hin1,Damour:2009vw,Bin-poi}.
Adopting such an effective action description of tidal effects, we will compute here several integrated tidal invariants associated with the worldlines of the two  members of a binary system undergoing hyperbolic motion. Our calculations will be performed within PM theory, at the first PM approximation level (1PM), {\it i.e.}, at first order in the gravitational constant $G$, but at all orders in velocities.
We will focus on quadratic and cubic invariants of both electric and magnetic types.

We will generally use units where $c=1$. 
The masses of the two gravitationally interacting bodies are denoted by $m_1$ and $m_2$. 
We then define the total rest mass of the system ($M$), the reduced mass ($\mu$) and the symmetric mass-ratio ($\nu$) as
\begin{eqnarray}
\label{eq:1.1}
M &\equiv& m_1+m_2\,,\qquad
\mu\equiv\frac{m_1 m_2}{M}\,,\nonumber\\
\nu &\equiv& \frac{\mu}{M}=\frac{m_1m_2}{(m_1+m_2)^2}\le \frac14
\,.
\end{eqnarray}
We will sometimes use the dimensionless mass ratios
\beq
\label{eq:1.2}
X_1\equiv \frac{m_1}{M} \,,\quad
X_2\equiv \frac{m_2}{M}= 1-X_1\,,
\eeq
with the link $\nu=X_1 X_2$.

\section{Perturbed on-shell action and scattering}

Here we shall follow Sec. II E of Ref.~\cite{Bini:2012gu} and show how some general properties of reduced actions allow one to simplify 
the discussion of additional effects  perturbing a basic dynamics.
We consider a two-body system whose interaction can be decomposed into some zeroth-order (unperturbed) dynamics modified by an additional interaction, of strength measured by a parameter $\epsilon$, say
\beq \label{S0S1}
S= S_0 + \epsilon \, S_1 \,.
\eeq
 When using an Hamiltonian formulation, 
such a  perturbed dynamics is described by a Hamiltonian of the general form,
\beq
\label{eq:1}
H({\mathbf p},{\mathbf q})=H_0({\mathbf p},{\mathbf q})+\epsilon \, H_1({\mathbf p},{\mathbf q})\,.
\eeq
Here $H_0({\mathbf p},{\mathbf q})$ describes the unperturbed dynamics while $\epsilon H_1({\mathbf p},{\mathbf q})$ describes a 
specific perturbation.
Examples of this very general setting are: i) free motion perturbed by the interaction mediated by some field; ii) geodesic motion in a black hole background perturbed by SF effects; iii) EOB description of 1PM gravity perturbed by higher PM interactions; iv) nonspinning dynamics perturbed by spin effects, etc.
In the following we consider the case where the tidal deformation of two interacting bodies perturbs the dynamics of two pointlike objects.

When studying, as we shall do, the relative dynamics of a two-body system considered in the center-of-mass (c.m.) frame, the phase-space variables reduce to that of one particle\footnote{Note, in passing, that this is the first element of the EOB approach to two-body dynamics.} of position ${\mathbf R}={\mathbf R}_1-{\mathbf R}_2$ and momentum ${\mathbf P}={\mathbf P}_1=-{\mathbf P}_2$. 
Then,  the energy conservation law yields
\beq
\label{eq:1}
E=H_0(R,P_R,P_\phi) +\epsilon H_1(R,P_R,P_\phi)\,,
\eeq
where $R=|{\mathbf R}|$.
Let us define $P_R^{(0)}(R;E,P_\phi)$ as the unperturbed solution of the energy conservation law, i.e., of the equation
\beq
E=H_0(R,P_R^{(0)},P_\phi)\,.
\eeq
Writing the solution of Eq. \eqref{eq:1} as 
\beq
P_R(R;E,P_\phi)=P_R^{(0)}(R;E,P_\phi)+\epsilon P_R^{(1)}(R;E,P_\phi)+O(\epsilon^2)
\eeq
 leads to the following first-order equation for $P_R^{(1)}$
\beq
\label{eq:2}
E=H_0(R,P_R^{(0)},P_\phi)+\frac{\partial H_0}{\partial P_R} \epsilon P_R^{(1)}+\epsilon H_1+O(\epsilon^2)\,,
\eeq
so that
\beq
\label{eq:3}
\epsilon P_R^{(1)}(R;E,P_\phi)=-\epsilon \left[\frac{H_1}{\frac{\partial H_0}{\partial P_R}}\right]_{P_R=P_R^{(0)}(R;E,P_\phi)}+O(\epsilon^2)\,.
\eeq
When considering bound motions, a crucial invariant quantity is the radial action, integrated over one radial period,
\beq
S_R(E,P_\phi)=\oint P_R(R;E,P_\phi) dR\,.
\eeq
As pointed out in Ref. \cite{Damour:2016gwp}, the analog of this invariant for scattering motion is the  (subtracted) radial action,
\beq
S_R^{\rm subt}(E,P_\phi)=\int_{-\infty}^{+\infty} dR [P_R(R; E,P_\phi)-P_R^{\rm free}(R; E,P_\phi)]\,,
\eeq
where $P_R^{\rm free}(R;E,P_\phi)$ denotes the value of $P_R$ in absence of any interaction~\footnote{Free motion here means motion in absence of any interaction, and not only of the additional interaction contained in $H_1$.}, and where the integral is taken over 
the full scattering motion [symbolically indicated by the time interval $ t \in ({-\infty}, {+\infty})$]. 
For instance, when using the (real) phase-space coordinates ${\mathbf R}={\mathbf R}_1-{\mathbf R}_2$ and  ${\mathbf P}={\mathbf P}_1=-{\mathbf P}_2$, one defines $P_R^{\rm free}(R; E,P_\phi)$ as the solution of 
\beq
E=\sqrt{m_1^2+{\mathbf P}^2}+\sqrt{m_2^2+{\mathbf P}^2}\,,
\eeq
with ${\mathbf P}^2=P_R^2+P_\phi^2/R^2$. The corresponding equation within the EOB formalism would be simply
\beq
\mu^2+{\mathbf P}_{\rm eob}^2=E_{\rm eff}^2\,,
\eeq
with ${\mathbf P}_{\rm eob}^2=(P_R^{\rm eob})^2+(P_\phi^{\rm eob})^2/R^2$,  $P_\phi^{\rm eob}=P_\phi$, and with
the EOB effective energy defined as  \cite{Buonanno:1998gg,Damour:2016gwp}
\beq
E_{\rm eff}=\frac{E^2-m_1^2-m_2^2}{2(m_1+m_2)}\,.
\eeq
The subtracted term $P_R^{\rm free}(R; E,P_\phi)$ has the effect both to render convergent~\footnote{Modulo the mild Coulomb 
logarithmic divergence when working in 3 space dimensions.}
 the radial action (which would otherwise diverge linearly at large $R$) and to subtract the free-motion contribution, $\pi$, from the scattering angle. Indeed, 
\beq
 -\int_{-\infty}^{+\infty} \frac{\partial}{\partial P_\phi} P_R{}^{\rm free} dR =\pi\,.
\eeq

The total angular change during scattering is
\beq
\Phi_{\rm scatt}= -\int_{-\infty}^\infty dR \frac{\partial}{\partial P_\phi} P_R(R; E, P_\phi)\,,
\eeq
where the integral is convergent because of the $P_\phi$ differentiation in the integrand.
The corresponding scattering angle $\chi\equiv\Phi_{\rm scatt}-\pi$ is then given by
\begin{eqnarray}
\chi&=& -\int_{-\infty}^\infty dR \frac{\partial}{\partial P_\phi} P_R(R; E, P_\phi)\\
&& +\int_{-\infty}^\infty dR \frac{\partial}{\partial P_\phi} P_R^{\rm free}(R; E, P_\phi)\nonumber\\
&=&  -\int_{-\infty}^\infty dR \frac{\partial}{\partial P_\phi} [P_R(R; E, P_\phi)-P_R^{\rm free}(R; E, P_\phi)]\,.\nonumber
\end{eqnarray} 
Therefore
\beq
\label{chi_sub_action}
\chi(E, P_\phi; \epsilon)=-\frac{\partial}{\partial P_\phi}S_R^{\rm subt}(E, P_\phi)\,,
\eeq
where the $P_\phi$-derivative could be taken out because the subtracted radial action is convergent for large $R$.

The scattering angle \eqref{chi_sub_action} is the full $\epsilon$-perturbed angle.
When expanding $\chi$ in series of $\epsilon$, 
\beq
\chi(E, P_\phi; \epsilon)=\chi^{(0)}(E, P_\phi)+\epsilon \chi^{(1)}(E, P_\phi)+O(\epsilon^2)\,,
\eeq
we find that 
\beq
\label{eq:chi0}
 \chi^{(0)}(E, P_\phi)=-\frac{\partial}{\partial P_\phi} \int_{-\infty}^\infty P_R^{(0)}(R; E, P_\phi)dR\,,
\eeq
and
\beq
\label{eq:chi1}
\epsilon \chi^{(1)}(E, P_\phi)=-\epsilon \frac{\partial}{\partial P_\phi} \int_{-\infty}^\infty P_R^{(1)}(R; E, P_\phi)dR\,.
\eeq
Inserting the expression of $P_R^{(1)}$, Eq. \eqref{eq:3}, in Eq. \eqref{eq:chi1} yields 
\beq
\epsilon \chi^{(1)}(E, P_\phi)=+\epsilon \frac{\partial}{\partial P_\phi} \int_{-\infty}^\infty \frac{H_1}{\frac{\partial H_0}{\partial P_R}}dR\,.
\eeq
According to Hamilton's equations $\frac{\partial H_0}{\partial P_R}$ can be replaced  by
the time derivative $\frac{dR}{dt}$ taken along the unperturbed, $H_0$-driven, motion so that 
\beq
\label{chi_1_ref}
\chi^{(1)}(E, P_\phi)=-\frac{\partial}{\partial P_\phi}  S_R^{(1)}  (E,P_\phi)\,.
\eeq
Here we have introduced the notation $S_R^{(1)}(E,P_\phi)$ for the $\epsilon$ piece of the subtracted radial action
\beq
S_R^{\rm subt}(E,P_\phi; \epsilon)= S_R^{\rm subt}(E,P_\phi; 0)+\epsilon S_R^{(1)}(E,P_\phi)+O(\epsilon^2)\,.
\eeq
From the above results, we can write the following explicit (equivalent) expressions for $S_R^{(1)} (E,P_\phi)$
\begin{eqnarray}
S_R^{(1)}(E,P_\phi)&=& \int_{-\infty}^{+\infty} P_R^{(1)} d R\nonumber\\
&=& -\int_{-\infty}^{+\infty} \frac{H_1}{\frac{\partial H_0}{\partial P_R}} dR \nonumber\\
&=& -\int_{-\infty}^{+\infty} dt^{(H_0)}H_1 \,.
\end{eqnarray}
In addition, using the general property, $\delta L({\mathbf q}, \dot{\mathbf  q},\ldots)  = - \delta H({\mathbf q},{\mathbf p})$,
 relating  a first-order change in a Lagrangian to a change in the corresponding Hamiltonian, we can directly relate $S_R^{(1)}(E,P_\phi)$
 to the $\epsilon$ piece of the original (Lagrangian-type) action, Eq. \eqref{S0S1}, namely
 \beq \label{S1onshell}
 S_R^{(1)}(E,P_\phi) = \left[ S_1 \right]^{{\rm on-}S_0{\rm -shell}}_{(E,P_\phi)}\,,
 \eeq
 where the notation on the right-hand side indicates that one must on-shell evaluate $S_1$ along a full 
$S_0$-driven (or $H_0$-driven) motion, with given total  energy and angular
 momentum. In the case where $\epsilon$ denotes the perturbation linked to the nonlocal tail effects in the orbital dynamics,  
this result was obtained in Ref. \cite{Bini:2017wfr}, where the gauge-invariant \lq\lq potential" for the perturbed scattering angle was denoted as 
$W^{(1)}(E,P_\phi)=-S_R^{(1)}(E,P_\phi)=+\int_{-\infty}^\infty dt^{(H_0)}H_1$.

The gauge-invariant nature of $S_R^{(1)}(E,P_\phi)$ allows one to easily transcribe its value within the EOB framework.
We recall that EOB theory formulates  the center-of-mass two-body dynamics in terms of a mass-shell constraint of the general form
\beq
\label{mass-shell}
g^{\mu\nu}_{\rm eff}P^{\rm eob}_\mu P^{\rm eob}_\nu +\mu^2 +Q=0\,,
\eeq
where $Q$ is a function in EOB phase-space (which is not simply quadratic in $P^{\rm eob}_\mu$).
When considering the $\epsilon$-perturbed version of the EOB mass-shell condition \eqref{mass-shell} one has the choice to parametrize perturbations either by modifying the effective metric $g^{\mu\nu}_{\rm eff}$ (when this is possible) or by changing  the $Q$  term in the mass-shell condition \eqref{mass-shell}, or by doing both. In several previous papers dealing with tidal effects   \cite{Damour:2009wj,Bini:2012gu} it was found convenient, when focusing on circular motions, to describe tidal effects by an additional term in $g^{\mu\nu}_{\rm eff}$ and more precisely in its main radial potential $A(R)=-g_{00}^{\rm eff}$. By contrast, here, as we are considering hyperbolic motions, it will be more convenient to
describe tidal effects by an additional (non-quadratic-in-momenta) term in $Q$.
Let us then consider a perturbed mass-shell of the form \eqref{mass-shell}, with a perturbed $Q$ of the general type
\beq
Q=Q^{(0)}+\epsilon Q^{(1)}\,.
\eeq 
For simplicity, we shall assume here that the unperturbed effective metric is spherically symmetric (as is the case
for non-spinning bodies):
\begin{eqnarray}
g_{\mu\nu}^{\rm eff}dx^\mu dx^\nu &=&-A(R_{\rm eob}) dt_{\rm eob}^2 +B(R_{\rm eob}) dR_{\rm eob}^2\nonumber\\
&& +R_{\rm eob}^2 (d\theta_{\rm eob}^2 +\sin^2 \theta _{\rm eob} d\phi^2_{\rm eob})\,.
\end{eqnarray}
The effective Hamiltonian, $H_{\rm eff}$, is obtained by solving the 
EOB mass-shell condition, Eq. \eqref{mass-shell}, with respect to $P_0^{\rm eob} = - H_{\rm eff}$, {\it i.e.},
\beq
H_{\rm eff}^2 =A(R_{\rm eob}) \left[ \frac{P_R^{2 \rm eob}}{B(R_{\rm eob})}+ \frac{P_\phi^{2 \rm eob}}{R^2_{\rm eob}}+\mu^2 +Q^{(0)}+\epsilon Q^{(1)}\right]\,.
\eeq
If we assume that $Q^{(0)}$ does not depend on $P_0^{\rm eob}$ (as, for example, was done in Ref. \cite{Damour2018}), we then find that
\beq
H^{\rm eff}(R_{\rm eob}, P_R^{\rm eob}, P_\phi^{\rm eob})=H_0^{\rm eff}+\epsilon H_1^{ \rm eff} + O(\epsilon^2)\,,
\eeq
with
\beq
\label{eff_H0}
\left(H_0^{ \rm eff}\right)^2 =A(R_{\rm eob}) \left[ \frac{P_R^{2 \rm eob}}{B(R_{\rm eob})}+ \frac{P_\phi^{2 \rm eob}}{R^2_{\rm eob}}+\mu^2 +Q^{(0)}\right]\,,
\eeq
and
\beq
\label{eff_H1}
H_1^{ \rm eff} =\frac{A}{2H_0^{\rm eff} }Q^{(1)}\,.
\eeq
Let us recall the crucial facts that --because of their gauge-invariant properties-- 
 both the EOB effective (subtracted) radial action, the total EOB angular momentum,  and the EOB scattering angle coincide with the corresponding \lq\lq real" 
physical quantities, 
\begin{eqnarray}
P_\phi^{\rm eob}&=& P_\phi^{\rm real} \,,\nonumber\\
S_R^{\rm eff,\, subt}(E_{\rm eff},P_\phi; \epsilon)&=&S_R^{\rm real,\, subt}(E_{\rm real},P_\phi; \epsilon)\,,\nonumber\\
\chi^{\rm eob}(E_{\rm eff},P_\phi; \epsilon)&=&\chi^{\rm real}(E_{\rm real},P_\phi; \epsilon)\,.
\end{eqnarray}
We also recall that the effective energy, $E_{\rm eff}=-P_0^{\rm eob}=H_{\rm eff}$ is related to the real energy 
$E_{\rm real}=H_{\rm real}$ by the energy map \cite{Buonanno:1998gg,Damour:2016gwp} 
\beq
H_{\rm real}=M\sqrt{1+2\nu\left(\frac{H_{\rm eff}}{\mu}-1 \right)}\,.
\eeq
One then easily finds that the $\epsilon$ piece of the  total effective (subtracted) radial action (which is equal to the real one) 
\begin{eqnarray}
S_R^{\rm eff,\, subt}(E_{\rm eff},P_\phi; \epsilon)&=& S_R^{\rm eff,\,subt}(E_{\rm eff},P_\phi; 0)\nonumber\\
&+&\epsilon S_R^{(1){\rm eff}}(E_{\rm eff},P_\phi)+O(\epsilon^2),\nonumber\\
\end{eqnarray}
  is equal to its real counterpart 
\beq
S_R^{(1){\rm eff}}(E_{\rm eff},P_\phi)=S_R^{(1){\rm real}}(E_{\rm real},P_\phi)\,,
\eeq
and is given by the following expressions
\begin{eqnarray}
\label{S_1_eff}
S_R^{(1){\rm eff}}(E_{\rm eff},P_\phi)&=& \int_{-\infty}^\infty P_R^{\rm eob}{}^{(1)} d R^{\rm eob}\nonumber\\
&=& -\int \frac{ dR^{\rm eob} }{\frac{\partial H^{\rm eff}_0}{\partial P_R^{\rm eob}}}  H_1^{\rm eff}\nonumber\\
&=&-
\int dt^{H_0^{\rm eff}}_{\rm eff} H_1^{\rm eff}\,.
\end{eqnarray}
Here $t^{H_0^{\rm eff}}_{\rm eff}$ denoted the effective time evolution parameter (defined by Hamilton's equations), evaluated along the unperturbed effective motion, so that   
\beq
\frac{ dR^{\rm eob} }{dt^{H_0^{\rm eff}}_{\rm eff}}=\frac{\partial H^{\rm eff}_0}{\partial P_R^{\rm eob}}\,.
\eeq
Substituting in Eq. \eqref{S_1_eff} the value of the perturbed effective Hamiltonian, Eq. \eqref{eff_H1}, also yields the expression
\beq
\label{eq_def_Q}
S_R^{(1){\rm eff}}=-\frac12 \int d\sigma_{(0)} Q^{(1)}\,,
\eeq
where
\beq
d\sigma_{(0)}=\frac{A dR^{\rm eob} }{  H^{\rm eff}_0 \frac{\partial H^{\rm eff}_0}{\partial P_{R}^{(0) {\rm eob}}  }}\,.
\eeq
In the cases where the $P_R$-dependence of the  unperturbed effective Hamiltonian is accurately described by 
\beq
(H^{\rm eff}_0)^2=A \left( \frac{(P_R^{\rm eob})^2}{B}+{P_R^{\rm eob}\hbox{-{\rm independent terms}}}\right)\,,
\eeq
the parameter $\sigma_0$ simplifies to 
\beq
d\sigma_{(0)}=
B \frac{dR^{\rm eob} }{P_{R}^{(0) {\rm eob}}}= \frac{dR^{\rm eob}}{P^{R}_{(0) {\rm eob}}}\,,
\eeq
where $P^{R}\equiv g_{\rm eff}^{RR}P_R=\frac{P_R}{B}$.
The corresponding formula for the perturbation of the scattering angle,
\beq
\chi^{(1)}= + \frac12 \frac{\partial}{\partial P_\phi} \int d \sigma_{(0)} Q^{(1)}\,,
\eeq
agrees with the result obtained in 
 Eq. (4.22) of  Ref. \cite{Damour2018}, where the unperturbed squared effective Hamiltonian was a Schwarzschild mass-shell condition
($g_{\rm eff}^{\mu\nu}=g_{\rm Schw}^{\mu\nu}$ and $Q^{(0)}=0$)  and where the $\epsilon$ parameter was $G^2$, with $Q^{(1)}=\sum_{k\geq 2} u^k q_k(H_{\rm eff}^{\rm Schw})$ (PM energy gauge). In that case $\sigma_{(0)}$ was the unperturbed ($\mu-$normalized) effective proper time along the geodesic Schwarzschild motion.

Summarizing: in a general $\epsilon$-perturbed situation, the crucial potential from which one can deduce scattering information is 
the on-shell  perturbed radial action $S_R^{(1)}(E,P_\phi) =  -\int_{-\infty}^{+\infty} dt^{(H_0)}H_1$, 
 integrated along the full hyperbolic motion. 
From the function $S_R^{(1)}(E,P_\phi)$ one then deduces, by $P_\phi$-differentiation, the scattering angle. Let us note in passing that
the $E$-differentiation of $S_R^{\rm subt}(E,P_\phi; \epsilon)$ yields the (full) Wigner time delay \cite{Wigner:1955zz}, so that $\partial S_R^{(1)}(E,P_\phi)/\partial E$ is the $O(\epsilon)$-correction to this time delay.
The time delay is a gauge-invariant observable quantity associated with a general scattering situation which has not received yet much attention in the gravitational physics literature. Let us also note, as recently pointed out in Ref. \cite{Damour:2019lcq}, that $S_R^{\rm subt}(E,P_\phi)$ is equal to the classical limit of the quantum phase shift (entering the partial wave decomposition of the quantum scattering amplitude)
\beq
S_R^{\rm subt}(E,P_\phi)=\frac{2\,\delta_l(E)}{\hbar} \,,
\eeq
where $l=P_\phi/\hbar$, see e.g., Eq. (4.5) of Ref.  \cite{Damour:2019lcq}.

\section{Worldline tidal actions}

In the following we will apply the general results of the previous section to the case where the unperturbed dynamics is that of two pointlike objects, and where the perturbation is due to the tidal deformations of two extended bodies, e.g., neutron stars. 

In an effective field theory description of $N$ extended (compact) objects, finite-size effects are treated by increasing the (leading-order) point-mass action \cite{Damour:1982wm}
\beq
\label{eq:2.1}
S_0=\int \frac{d^4 x}{16 \pi G}\, \sqrt{-g} R-\sum_{A=1}^N \int m_A  d\tau_A\,,
\eeq
by additional, nonminimal, couplings involving higher-order derivatives of the field evaluated along the worldlines of the bodies 
\cite{Damour:1995kt,Damour:1998jk,Goldberger:2004jt,porto06,porto06prl,porto08,Levi:2014gsa,Levi:2015msa}. 
Here $d\tau_A= \sqrt{- g^A_{\mu \nu} d z_A^\mu d z_A^\nu}$ is the (regularized) proper time along the worldline $z_A^\mu(\tau_A)$ of body $A$, with $4$-velocity $u_A^\mu = d z_A^\mu/d\tau_A$. In the matched-asymptotic-expansion approach to
$N$-body dynamics,  body $A$ feels the gravitational field of the whole interacting $N$-body system via a smooth \lq\lq external metric'' $G^A_{\alpha \beta}(X^\gamma)$ defined in the local coordinate system $X^\alpha_A$ attached to body $A$ \cite{Damour:1982wm,Zhang86,Damour:1990pi,Damour:1991yw,Damour:1992qi,Damour:1993zn}. 
Non-minimal couplings are expressed in terms of two types of tidal tensors computed from the latter external metric~\footnote{In practice, the
use of dimensional regularization allows one to compute these tensors, and their invariants, directly from the singular $N$-body metric 
$g_{\mu \nu}(x^\lambda)$.}: the gravitoelectric $G_L^A(\tau_A)\equiv G^A_{a_1\ldots a_l}(\tau_A)$, 
and gravitomagnetic $H_L^A(\tau_A)\equiv H^A_{a_1\ldots a_l}(\tau_A)$, 
tidal tensors, together with their proper time derivatives (we follow here the normalization and notation of Refs. 
\cite{Damour:1990pi,Damour:1991yw,Damour:2009wj}). These tensors are symmetric, tracefree, and spatial with respect to $u_A$.  [The spatial indices $a_i=1,2,3$ refer to the local body-fixed coordinates $X^\alpha_A=(\tau_A, X_A^a)$, attached to body $A$.] In terms of these tidal tensors, the general nonminimal worldline action (for nonspinning bodies) has the form \cite{Damour:1995kt,Damour:1998jk,Goldberger:2004jt,Damour:2009wj}
\begin{eqnarray}
\label{S_non_min_eq}
S_{\rm nonmin} 
&=&\sum_{A}S_{\rm nonmin} ^{A}=\sum_{A}\sum_{l \ge 2} S_{\rm nonmin} ^{A,l}\,,
\end{eqnarray}
with $S_{\rm nonmin} ^{A,l}$ given by
\begin{widetext}
\begin{eqnarray}
\label{Snonmin}
S_{\rm nonmin} ^{A,l}&=& \frac12 \frac{1}{l!}\left[\mu_A^{(l)}\int d\tau_A (G_L^A(\tau_A))^2
+\frac{l}{l+1}\sigma_A^{(l)}\int d\tau_A (H_L^A(\tau_A))^2\right. \nonumber\\
&& \left. +\mu'_A{}^{(l)}\int d\tau_A (\dot G_L^A(\tau_A))^2
+\frac{l}{l+1}\sigma'_A{}^{(l)} \int d\tau_A (\dot H_L^A(\tau_A))^2
+\ldots
\right]
\end{eqnarray}
\end{widetext}
where $L=i_1\ldots i_l$ is a multi-index and $\mu_A^{(l)}$, $\sigma_A^{(l)}$, $\mu_A^{(l)}$, $\mu'^{(l)}_A$ are (electric or magnetic type) tidal coefficients. Moreover,  $\dot G_L^A(\tau_A)=dG_L^A/d\tau_A$ and the ellipsis refer  either to higher-order proper-time derivatives of $G_L^A$ and $H_L^A$, or to higher-than-quadratic invariant monomials, built with   
$G_L^A$ and $H_L^A$ and their proper-time derivatives. 
The allowed monomials are restricted by certain symmetry constraints. For example, in the non-spinning case where the dynamics should be symmetric under time and space reflections, monomials which imply time reversal or space reversal are not allowed.  See Ref. \cite{Bini:2014zxa} for additional details.

Let us define the \lq\lq electric" and \lq\lq magnetic" parts of the Riemann tensor evaluated along the worldline ${\mathcal L}_A$
of body $A$ (the star denoting the dual: 
$R^*_{\alpha\beta\mu\nu} \equiv \frac12 \eta_{\alpha\beta\gamma\delta}{R^{\gamma\delta}}_{\mu\nu}$
with $\eta_{0123}= +\sqrt{-g})$) 
\begin{eqnarray}
\label{riemann_em}
E^A_{\alpha\beta}(u_A)&\equiv& R^A_{\alpha\mu\beta\nu}(x) u_A^\mu u_A^\nu\,,\nonumber\\
B^A_{\alpha\beta}(u_A)&\equiv&[R^*]^A_{\alpha\mu\beta\nu}(x) u_A^\mu u_A^\nu\,,
\end{eqnarray}
and their three-index (octupolar) counterparts (with the $u_A$-orthogonal projector $P(u_A)\equiv g+u_A\otimes u_A$)
\begin{eqnarray}
\tilde E^A_{\alpha\beta\gamma}(u_A)&\equiv&{\rm Sym}_{\alpha\beta\gamma}\,(P(u_A)_\alpha^\mu \nabla_\mu R_{\beta \rho \gamma \nu})u_A^\rho u_A^\nu\,, \nonumber\\
\tilde B^A_{\alpha\beta\gamma}(u_A)&=&{\rm Sym}_{\alpha\beta\gamma}\,(P(u_A)_\alpha^\mu \nabla_\mu R^*_{\beta \rho \gamma \nu})u_A^\rho u_A^\nu\,.
\end{eqnarray}
[We added a tilde over them for clarity when later considering their squares.]
We will also consider the invariants constructed with 
\begin{eqnarray}
\dot E^A(u_A)_{\alpha\beta}&=&\nabla_{u_A} E(u_A)_{\alpha\beta}\,,\nonumber\\
\dot B^A(u_A)_{\alpha\beta}&=&\nabla_{u_A} B(u_A)_{\alpha\beta}\,.
\end{eqnarray}
In the normalization used to define the nonminimal action \eqref{Snonmin},
the corresponding electric-type and magnetic-type tidal quadrupolar and octupolar tensors, $G_{ab}$, $H_{ab}$, $G_{abc}$,  $H_{abc}$ read
\begin{eqnarray}
\label{eq:2.4}
G^A_{\alpha\beta} &\equiv& -E^A_{\alpha\beta}(u_A)\,, \nonumber\\
H^A_{\alpha\beta} &\equiv & 2\,  B^A_{\alpha\beta}(u_A)\,, \nonumber\\
G^A_{\alpha\beta\gamma}&\equiv&-\tilde E^A_{\alpha\beta\gamma}(u_A)\,, \nonumber\\
H^A_{\alpha\beta\gamma}&\equiv&2 \, \tilde B^A_{\alpha\beta\gamma}(u_A)\,.
\end{eqnarray}
These expressions use the fact that, at the approximation order where we work here, the Ricci tensor vanishes so that the various defined tensors are symmetric and  {\it tracefree}.

The first few terms of the above expansion of the nonminimal worldline action are 
\begin{eqnarray}
\label{eq:2.2}
S_{G_{ab}}&=& \frac{1}{4} \, \mu_A^{(2)} \int d\tau_A \, G_{\alpha\beta}^A \, G_A^{\alpha\beta}\,,\nonumber\\
S_{H_{ab}}&=& \frac{1}{6 } \, \sigma_A^{(2)} \int d\tau_A \, H_{\alpha\beta}^A \, H_A^{\alpha\beta}\,,\nonumber\\
S_{G_{abc}}&=& \frac{1}{12} \, \mu_A^{(3)} \int d\tau_A \, G_{\alpha\beta\gamma}^A \, G_A^{\alpha\beta\gamma}\,,\nonumber\\
S_{\dot G_{ab}}&=& \frac{1}{4 } \, \mu'^{(2)}_A \int d\tau_A \dot G_{\alpha\beta}^A\dot G_A^{\alpha\beta}\,.
\end{eqnarray}
Here $\mu_A^{(2)}$, $\sigma_A^{(2)}$, $\mu_A^{(3)}$, $\mu'^{(2)}_A$ are tidal coefficients. See, {\it e.g.}, Ref. \cite{Damour:2009vw}
for their computation, and their links with corresponding dimensionless Love numbers.
Higher-order invariants involve higher-than-quadratic tidal scalars, e.g.,
cubic in $G_{ab}^A$ 
\beq
\label{eq:2.3}
\int d\tau_A G^A{}_{ab}G^A{}^{bc}G^A{}_{c}{}^{a}\,.
\eeq
Below, we will explicitly consider only the  tidal invariants associated with the body labeled 1 (with mass $m_1$) in a binary system 
(i.e., $N=2$). The other tidal invariants are then simply obtained by a $1\leftrightarrow 2$ exchange.
We shall  sometimes suppress the body label $A=1$.

The quadrupolar electric-type tidal tensor (\ref{eq:2.4}), in non-spinning comparable mass binary systems,  has been computed to 1PN fractional accuracy in Refs. \cite{Damour:1992qi,Damour:1993zn} (see also Refs. \cite{Vines:2010ca,Taylor:2008xy} fore more details). Ref. \cite{JohnsonMcDaniel:2009dq} has also computed to 1PN accuracy the octupolar electric-type tidal tensor, $G_{abc}$, and the quadrupolar magnetic-type tidal tensor $H_{ab}$. The significantly more involved calculation of tidal effects (along general orbits, but still in the case of non-spinning binary systems) at the 2PN fractional accuracy has been done in Ref. \cite{Bini:2012gu}.

In the present work we will evaluate  the tidal invariants
\begin{eqnarray}
&& E(u_1)^2\equiv E(u_1)_{\alpha\beta}E(u_1)^{\alpha\beta}\,,\nonumber\\
&& B(u_1)^2\equiv B(u_1)_{\alpha\beta}B(u_1)^{\alpha\beta}\,,\nonumber\\
&& E(u_1)^3\equiv E(u_1)_{\alpha\beta}E(u_1)^{\beta\gamma}{E(u_1)_{\gamma}}^{\alpha}\,,\nonumber\\
&& B(u_1)^3\equiv B(u_1)_{\alpha\beta}B(u_1)^{\beta\gamma}{B(u_1)_{\gamma}}^{\alpha}\,, \nonumber\\
&& \tilde E(u_1)^2  \equiv \tilde E(u_1)_{\alpha\beta\gamma} \tilde E(u_1)^{\alpha\beta\gamma}  \,,\nonumber\\
&&  \tilde B(u_1)^2   \equiv \tilde B(u_1)_{\alpha\beta\gamma} \tilde B(u_1)^{\alpha\beta\gamma}   \,,\nonumber\\
&&  \dot E(u_1)^2   \equiv \dot E(u_1)_{\alpha\beta} \dot E(u_1)^{\alpha\beta} \,,\nonumber\\
&&  \dot B(u_1)^2 \equiv \dot B(u_1)_{\alpha\beta} \dot B(u_1)^{\alpha\beta} \,,
\end{eqnarray}
along the worldline ${\mathcal L}_1$ of the mass $m_1$ using  PM theory, at the first PM approximation level, i.e., limiting 
ourselves to the first-order in $G$ but including all orders in $v/c$. [The tidal coefficients $\mu_A^{(l)}$, etc., contain a factor 
$1/G$, so that the $O(G)$ tidal action is obtained by inserting the linearized gravity tidal tensors in the nonminimal worldline action
\eqref{Snonmin}.]

As explained in Sec. II, the integrated values of all the above quantities along the worldline ${\mathcal L}_1$ define gauge-invariant quantities of direct dynamical significance interesting for describing tidal effects in scattering situations.

\section{The first Post-Minkowskian approximation}

As proven in Sec. II above, see Eq. \eqref{S1onshell}, the perturbed radial action $S_R^{(1)}(E, P_{\phi})$ is simply equal to the 
on-shell value of the additional worldline action associated with tidal effects,
taken along an unperturbed motion with given values of energy and angular momentum.
In other words, we wish to integrate the tidal action $S_{\rm nonmin}$, Eq. \eqref{Snonmin}, along an unperturbed hyperbolic two-body motion with
given energy and angular momentum. This computation can in principle be done to all orders of post-Minkowskian gravity,
which would mean taking into account all powers of $G$ in the gravitational interaction of the two bodies.
In the present paper we will solve this problem at the lowest order, where the interbody gravitational field entering the tidal action will be obtained by solving the linearized Einstein's equations. At this leading PM order, when evaluating $S_{\rm nonmin}$,
the worldlines of the two bodies can be treated as free motion worldlines (i.e., straight lines in Minkowski spacetime).

At the 1PM order, i.e. when solving the linearized Einstein equations in harmonic coordinates, 
the metric generated by our binary system is of the form $g_{\mu \nu} = \eta_{\mu \nu} + h_{\mu \nu} +O(G^2)$,
with 
\beq
h_{\mu \nu}= h_{1 \, \mu \nu} + h_{2 \,\mu \nu}\,,
\eeq
where $ h_{1 \, \mu \nu}$ is generated by ${\mathcal L}_1$ and $ h_{2 \, \mu \nu}$  by ${\mathcal L}_2$. 
When computing the tidal effects 
along ${\mathcal L}_1$, the external metric is simply the contribution
$ h_{2 \, \mu \nu}$  from ${\mathcal L}_2$, 
containing  a factor $Gm_2$. Finally, we deal along ${\mathcal L}_1$
with the (regular) metric
\beq
\label{metrica_tot}
g_{1 \,\mu \nu} = \eta_{\mu \nu} + h_{2 \, \mu \nu} \,,
\eeq
 with   $ h_{2 \, \mu \nu} \propto G m_2$.   
It is straightforward to compute the 1PM-accurate tidal tensors from the  
1PM metric \eqref{metrica_tot} generated by ${\mathcal L}_2$. Using, for instance, the results given in Appendices A and B of Ref. \cite{Bel:1981be},
and using the simplifying fact that, at this order, we can consider that ${\mathcal L}_2$ is a straight worldline (with tangent $u_2$),
we have (at an arbitrary field point $x^\mu$)
\beq
h_{2 \,\mu \nu}(x) = 2 \frac{G m_2}{R_2} \left( 2 \, u_{2 \, \mu} u_{2 \, \nu} + \eta_{\mu\nu} \right)\,,
\eeq
which is conveniently rewritten in the form
\begin{eqnarray} 
\label{h2}
h_{2 \,\mu \nu}(x) &=& \Phi(x) H_2{}_{\mu\nu}\,,\nonumber\\
\Phi(x)&\equiv&  \frac{2 G m_2}{R_2(x)} \,, \nonumber\\
 H_2{}_{\mu\nu}&\equiv&\eta_{\mu\nu}+ 2 \, u_{2 \, \mu} u_{2 \, \nu}\,.  
\end{eqnarray}
Here $R_2=R_2(x)$ denotes the Poincar\'e-invariant orthogonal distance between the field point $x$ and the straight worldline 
${\mathcal L}_2$. Explicitly, $R_2(x) = | x - z_2^\perp(x)|$, is the modulus of the four-vector
\beq
R_2^\mu(x) =  x^\mu - {z_2^\mu}_\perp(x)\,,
\eeq
where ${z_2^\mu}_\perp(x)$ denotes the foot of the perpendicular of the field point $x$ on the line ${\mathcal L}_2$. 

The expressions of the two (straight) worldlines are the following
\begin{eqnarray}
z_1(\tau_1)&=& z_1(0)+ u_1 \tau_1 + O(G)\,,\nonumber\\
z_2(\tau_2)&=& z_2(0)+ u_2 \tau_2 + O(G)\,,
\end{eqnarray}
with $u_1$ and $u_2$ constant  vectors.
[In the case where one must take into account the $O(G)$ curvature of ${\mathcal L}_2$ the expression of 
$h_{2 \,\mu \nu}(x)$ should involve the half-sum of  retarded and advanced tensor potentials generated by ${\mathcal L}_2$.]
It is convenient to choose the origins of the proper-time parameters along ${\mathcal L}_1$ and ${\mathcal L}_2$ such that
the corresponding connecting four-vector
\beq
 b^\mu \equiv z_1^\mu(0) -  z_2^\mu(0)\,,
\eeq
is  perpendicular  {\it both} to $u_1$ and to $u_2$. The vector $b^\mu$ can be thought of as being the 
 Poincar\'e-invariant four-vectorial impact parameter of ${\mathcal L}_1$ with respect to  ${\mathcal L}_2$, corresponding
 to a moment of closest approach of the two bodies.
[The vectorial impact parameter of ${\mathcal L}_2$ with respect to ${\mathcal L}_1$ is simply $ z_2^\mu(0) -  z_1^\mu(0) = - b^\mu$.]

As we are computing spacetime scalars, we can choose a coordinate system which simplifies our computations.
We find convenient to use coordinates adapted to ${\mathcal L}_1$, i.e., coordinates with respect to which the first body is at rest,
so that (see Appendix A for more details)
\beq
\tau_1 = t, \qquad u_1=\partial_t, \qquad z_1(t)=t\partial_t +  b \partial_x \,.
\eeq
The body 2 is then moving with respect to these coordinates and we assume that it is in motion along the (negative) $y$ axis, with
\beq
u_2=\gamma \partial_t -\sqrt{\gamma^2-1}\partial_y, \qquad z_2(\tau_2)= \tau_2\,u_2 \,.
\eeq
The Lorentz gamma factor between the two worldlines,
\beq
\gamma \equiv - u_1 \cdot u_2\,,
\eeq
will play an important role in all the formulas below.
[Here, and below, the scalar product $u \cdot v$ is the Poincar\'e-Minkowski one.]

We have assumed that $b^\mu$ is along the $x$-axis, and that
\beq
z_1(0)=b \partial_x\,,\qquad   z_2(0)=0\,.
\eeq
In this way  we have, for example, that 
\beq
z_2^\perp (x)=\tau_2^\perp (x) \,u_2 \,,
\eeq
where $\tau_2^\perp(x)$ is identified by the condition
\beq
(x-z_2^\perp (x))\cdot u_2=0\,,
\eeq
namely
\beq
u_2\cdot (x-\tau_2^\perp \, u_2 )=0\,,
\eeq
and hence 
\beq
\tau_2^\perp(x)=-u_2\cdot x\,.
\eeq
Consequently
\begin{eqnarray}
z_2^\perp (x)&=&-u_2 (u_2\cdot x)=T(u_2)x\,,\nonumber\\ 
x-z_2^\perp (x)&=&P(u_2)x\,,
\end{eqnarray}
where $T(u_2)=-u_2\otimes u_2$ projects along $u_2$ while $P(u_2)=I-T(u_2)=I+u_2\otimes u_2$ projects orthogonally to $u_2$.
Explicitly, with respect to the chosen coordinate system,
\begin{eqnarray}
z_2^\perp (x)&=& (\gamma t +\sqrt{\gamma^2-1} \, y) u_2 \,,\nonumber\\
x-z_2^\perp (x)&=& x^\alpha\partial_\alpha -(\gamma t +\sqrt{\gamma^2-1} \, y) u_2\,.
\end{eqnarray}
Moreover,
\begin{eqnarray}
\label{R2_def_expl}
R_2(x)^2&=&[x-z_2^\perp (x)]^2=x^2+(x\cdot u_2)^2\\
&=& (\gamma^2-1)t^2+x^2+\gamma^2 y^2 +z^2+2\gamma \sqrt{\gamma^2-1} t y\,, \nonumber\\
\end{eqnarray}
so that
\begin{eqnarray}
R_2(x)&=&|P(u_2)x| \nonumber\\
&=&\sqrt{ (\gamma^2-1)t^2+x^2+\gamma^2 y^2 +z^2+2\gamma \sqrt{\gamma^2-1} t y}\,. \nonumber\\
\end{eqnarray}
The corresponding perpendicular distance between the field point $x$ and ${\mathcal L}_1$ reads
\beq
R_1(x)=|P(u_1)x|=\sqrt{x^2+y^2+z^2}\,.
\eeq
One can also evaluate the four-vector $R_2^\mu(z_1(\tau_1))$ connecting a point of ${\mathcal L}_1$ to its orthogonal foot on 
${\mathcal L}_2$, i.e.
\begin{eqnarray} 
z_1(\tau_1) - z_{2}^\perp\left(z_1(\tau_1)\right)&=& P(u_2) z_1(\tau_1) \nonumber\\
&=& P(u_2)(b\partial_x + \tau_1 u_1)\nonumber\\
&=& b\partial_x +  \tau_1 P(u_2)u_1  \,.
\end{eqnarray}
The length of the latter (spacelike) vector reaches its minimum value $b$ when $\tau_1=0$.

One can also easily evaluate the retarded and advanced points on ${\mathcal L}_1$ and ${\mathcal L}_2 $
associated with the spacetime point $x$, say $z_{1R,A}=z_1(\tau_{1R,A})$ and $z_{2R,A}=z_2(\tau_{2R,A})$.
They correspond to the two roots of the null conditions
\beq
(x-z_{1R,A})^2=0, \; (x-z_{2R,A})^2=0\,.
\eeq
These roots are respectively given by 
\begin{eqnarray}
\tau_{1R,A}&=&-(x\cdot u_1)\pm \sqrt{(x\cdot u_1)^2+x^2}\nonumber\\
&=&-(x\cdot u_1)\pm R_1(x) \nonumber\\
&=& t \pm R_1(x)\,,
\end{eqnarray}
and
\begin{eqnarray}
\tau_{2R,A} &=& -(x\cdot u_2)\pm \sqrt{(x\cdot u_2)^2+x^2}\nonumber\\
&=&-(x\cdot u_2)\pm R_2(x)\,.
\end{eqnarray}

Well known dynamical quantities at the first post-Minkowskian approximation level, determined in previous work,  are the scattering angle $\chi$ and the spin holonomy $\theta$.
The first quantity, $\chi$, defines the direction of the final momentum in the center-of-mass frame after the full scattering process, while the second one, $\theta$, defines the precession angle of a spin vector (for spinning bodies) again after the full scattering process (it is known today up to the 2PM level included \cite{Bini:2018ywr}). Their 1PM expressions are the following \cite{Bini:2017xzy}
\begin{eqnarray}
\chi&=& \frac{2GM(2\gamma^2-1)}{b(\gamma^2-1)}+O(G^2)\,,\nonumber\\ 
\theta &=&\frac{2GM}{b (\gamma^2-1)}\left\{\frac{m_1}{M} (2\gamma^2-1)(h-1)\right.\nonumber\\ 
&&\left. +\frac{m_2}{M}[(2\gamma^2-1)h-\gamma]\right\}+O(G^2)\,,
\end{eqnarray}
where
\beq
h(\gamma, \nu)\equiv\sqrt{1+2\nu(\gamma-1)}
=\frac{E}{M}\,,
\eeq
where $E$ is the total c.m. energy (asymptotically given by $E=\sqrt{m_1^2+{\mathbf P}^2}+\sqrt{m_2^2+{\mathbf P}^2}$).
Let us now extend the list of gauge-invariant scattering observables by computing integrated tidal scalars.

We must evaluate along ${\mathcal L}_1$  the combinations of partial derivatives of  $h_{2 \,\mu \nu}(x)$ 
entering the tidal tensor expressions, and then integrate them over ${\mathcal L}_1$,  using e.g., $dx^{\lambda} = u_1^{\lambda} d \tau_1$. 

The partial derivatives with respect to $x^\alpha$ of $h_{2 \,\mu \nu}(x)$ (using again $\dot u_2=O(G)$) are given by
\begin{eqnarray} 
\label{Dh2}
\partial_\alpha h_{2 \,\mu \nu}(x)&=& \Phi_{,\alpha} H_2{}_{\mu\nu}+O(G^2)\,,\nonumber\\
\partial_{\alpha \beta} h_{2 \,\mu \nu}(x)&=& \Phi_{,\alpha \beta} H_2{}_{\mu\nu}+O(G^2)\,,
\end{eqnarray}
where (with $R_{2\alpha}=\eta_{\alpha\beta}R_2^\beta$)
\begin{eqnarray}
\Phi_{,\alpha}&=&-2\frac{Gm_2}{R_2^3} R_{2\alpha}\,,\nonumber\\
\Phi_{,\alpha\beta}&=&\frac{6Gm_2}{R_2^5}\left(R_{2\alpha}R_{2\beta}-\frac13 R_2^2 P(u_2)_{\alpha\beta}\right)\,.
\end{eqnarray}
After differentiation one must replace $x\to x_1(\tau_1)$.
 
For example, the Riemann tensor components are given by
\begin{eqnarray}
R_{\alpha\beta\gamma\delta}&=&
H_2{}_{\alpha [\delta}\Phi_{,\gamma ]\beta}-H_2{}_{\beta [\delta}\Phi_{,\gamma ]\alpha}
\,.
\end{eqnarray}
When writing down the explicit expressions of the electric and magnetic components of the
Riemann tensor along $u_1$, $E(u_1)_{\alpha\beta}$, and $B(u_1)_{\alpha\beta}$, 
it is useful to define the following {\it past}-directed timelike vector
\beq
V_2^\mu \equiv 2 (u_1 \cdot u_2) \, u_2^\mu + u_1^\mu \,, \qquad  V_2 \cdot V_2 = -1 \,.
\eeq
Note that $V_2$ is  {\it asymmetric} under the $1 \leftrightarrow 2$ exchange.  

We find
\begin{eqnarray}
E(u_1)_{\alpha\gamma}&=&\frac12 \left(
 V_{2\alpha} u_1^\beta \Phi_{,\beta\gamma}
-(u_1\cdot V_2)  \Phi_{,\alpha\gamma}\right. \nonumber\\
&& \left.-H_2{}_{\alpha\gamma}u_1^\beta u_1^\delta \Phi_{,\beta\delta}
+V_{2\gamma}u_1^\delta\Phi_{,\alpha\delta}
 \right)\,,\nonumber\\
B(u_1)^{\alpha}{}_{\gamma}&=&\frac12 u_{1\, \mu}\eta^{\mu \alpha\rho \sigma} \left[ V_{2\rho}  \Phi_{,\sigma\gamma}-H_2{}_{\gamma\rho}\Phi_{\sigma\delta}u_1^\delta \right]\,,
\end{eqnarray}
where
\begin{eqnarray}
H_2{}_{\alpha\delta} u_1^\delta &=&V_{2\alpha}\,,\nonumber\\ 
H_2{}_{\alpha\delta} u_1^\alpha u_1^\delta &=&u_1\cdot V_2=-1+2(u_1\cdot u_2)^2\,.
\end{eqnarray}
With our choice of coordinates, we have $u_1\cdot u_2=-\gamma$, $u_1\cdot V_2=2\gamma^2-1$, and
\begin{eqnarray}
E(u_1)_{\alpha\gamma}&=&\frac12 \left[
 V_{2\alpha} \Phi_{,0\gamma}
-(2\gamma^2-1)  \Phi_{,\alpha\gamma}\right.\nonumber\\
&& \left. -H_2{}_{\alpha\gamma} \Phi_{,00}
+V_{2\gamma} \Phi_{,\alpha 0}
 \right]\,,\nonumber\\
B(u_1)^{\alpha}{}_{\gamma}&=&\frac12 \epsilon^{\alpha\rho \sigma} \left[ V_{2\rho}  \Phi_{,\sigma\gamma}-H_2{}_{\gamma\rho}\Phi_{\sigma 0} \right]\,,
\end{eqnarray}
where $\epsilon^{\alpha\rho \sigma}\equiv u_{1\, \mu}\eta^{\mu \alpha\rho \sigma}$. 

We first evaluate the values of several scalar tidal functions of the proper time ($\tau_1=t$),   along the worldline ${\mathcal L}_1$:
$ f(t) \equiv f(x(t))$.
For example, the instantaneous value of the invariant $E(u_1)^2$ at proper time $\tau_1=t$ is given by
\begin{eqnarray}
E(u_1)^2&=& \frac{18 G^2 m_2^2  }{[(\gamma^2-1)t^2+b^2]^5}\left[\frac13 (\gamma^2-1)^2 t^4\right. \nonumber\\
&+&\left(\gamma^4-\frac43 \gamma^2+\frac13\right) b^2 t^2\nonumber\\
&+& \left.\left(\gamma^4-\gamma^2+\frac13\right)b^4\right]\,.
\end{eqnarray}
It is then convenient to introduce the following shorthand  notation for the corresponding full proper-time integral
\beq
\langle f \rangle \equiv\int_{-\infty}^\infty f(x(\tau_1)) d\tau_1\,.
\eeq
Our final results have the form
\begin{eqnarray}
\langle  E(u_1)^2 \rangle &=& \frac{G^2 m_2^2}{b^5}{\mathcal F}_{E^2}(\gamma)\,,\nonumber\\
\langle  B(u_1)^2 \rangle &=& \frac{G^2 m_2^2}{b^5}{\mathcal F}_{B^2}(\gamma)\,,\nonumber\\
\langle  E(u_1)^3 \rangle &=& \frac{G^3 m_2^3}{b^8}{\mathcal F}_{E^3}(\gamma)\,,\nonumber\\
\langle  B(u_1)^3 \rangle &=& \frac{G^3 m_2^3}{b^8}{\mathcal F}_{B^3}(\gamma)\,,\nonumber\\
\langle  \tilde E(u_1)^2 \rangle &=& \frac{G^2 m_2^2}{b^7}{\mathcal F}_{\tilde E^2}(\gamma)\,,\nonumber\\
\langle  \tilde B(u_1)^2 \rangle &=& \frac{G^2 m_2^2}{b^7}{\mathcal F}_{\tilde B^2}(\gamma)\,,\nonumber\\
\langle  \dot E(u_1)^2 \rangle &=& \frac{G^2 m_2^2}{b^7}{\mathcal F}_{\dot E^2}(\gamma)\,,\nonumber\\
\langle  \dot B(u_1)^2 \rangle &=& \frac{G^2 m_2^2}{b^7}{\mathcal F}_{\dot B^2}(\gamma)\,,
\end{eqnarray}
where we have separated  scaling prefactors from functions giving the dependence of the various quantities on the Lorentz factor $\gamma$.
Defining
\beq
p_\infty\equiv \sqrt{\gamma^2-1}\,,
\eeq
the various functions ${\mathcal F}_{X}(\gamma)$ are given by
\begin{eqnarray}
\label{eq_all_tidals}
{\mathcal F}_{E^2}(\gamma)&=&  \frac{9 \pi(35\gamma^4-30\gamma^2+11)}{64\sqrt{\gamma^2-1}} \,,\nonumber\\
&=& \frac{9 \pi(16+ 40p_\infty^2+35p_\infty^4)}{64p_\infty }\,, \nonumber\\
{\mathcal F}_{B^2}(\gamma)&=& \frac{45\pi \sqrt{\gamma^2-1}(1+7\gamma^2)}{64}\,,\nonumber\\
&=&  \frac{45\pi p_\infty(8+7 p_\infty^2)}{64}\,,\nonumber\\
{\mathcal F}_{E^3}(\gamma)&=& -\frac{192(40\gamma^4-36\gamma^2 +7)}{385\sqrt{\gamma^2-1}}\,,\nonumber\\
&=& -\frac{192(11+44 p_\infty^2 +40 p_\infty^4)}{385p_\infty}\,, \nonumber\\
{\mathcal F}_{B^3}(\gamma)&=&0\,,\nonumber\\
{\mathcal F}_{\tilde E^2}(\gamma)&=&  \frac{75\pi (21\gamma^6+385 \gamma^4-305\gamma^2+91)}{512 \sqrt{\gamma^2-1}}\,,\nonumber\\
&=&  \frac{75\pi (192+528p_\infty^2+448 p_\infty^4+21 p_\infty^6)}{512 p_\infty}\,,\nonumber\\
{\mathcal F}_{\tilde B^2}(\gamma)&=& \frac{525 \pi\sqrt{\gamma^2-1}(3\gamma^4+58 \gamma^2+3)}{512}\,,\nonumber\\
&=&  \frac{525 \pi p_\infty(64+64 p_\infty^2+3p_\infty^4)}{512}\,,\nonumber\\
{\mathcal F}_{\dot E^2}(\gamma)&=&  \frac{225\pi \sqrt{\gamma^2-1} (21\gamma^4-14\gamma^2+9)}{512} \,,\nonumber\\
&=& \frac{225\pi p_\infty (16+28 p_\infty^2+21 p_\infty^4)}{512}\,, \nonumber\\
{\mathcal F}_{\dot B^2}(\gamma)&=& \frac{1575 \pi (\gamma^2-1)^{3/2}(3\gamma^2+1)}{512}\,,\nonumber\\
&=& \frac{1575 \pi p_\infty^3(3p_\infty^2+4)}{512}\,.
\end{eqnarray}

The high energy (HE, i.e., $\gamma \to \infty$) behaviors of those functions are:
\begin{eqnarray}
\label{eq_all_tidals_HE}
{\mathcal F}_{E^2}(\gamma)&\overset{\rm HE}{=}{\mathcal F}_{B^2}(\gamma)\overset{\rm HE}{=} & \frac{315}{64}\pi\gamma^3 \,,\nonumber\\
{\mathcal F}_{\tilde E^2}(\gamma)&\overset{\rm HE}{=}{\mathcal F}_{\tilde B^2}(\gamma)\overset{\rm HE}{=}&\frac{1575}{512}\pi\gamma^5 \,,\nonumber\\
{\mathcal F}_{\dot E^2}(\gamma)&\overset{\rm HE}{=}{\mathcal F}_{\dot B^2}(\gamma)\overset{\rm HE}{=}&\frac{4725}{512}\pi\gamma^5   \,,
\end{eqnarray}
while
\beq
{\mathcal F}_{E^3}(\gamma)\overset{\rm HE}{=}  -\frac{1536}{77}\gamma^3\,.
\eeq
On the other hand,  ${\mathcal F}_{B^3}(\gamma)\equiv 0$ at all energies because of its time-reversal antisymmetry.

Note the  fact, visible on Eq. \eqref{eq_all_tidals_HE}, that the high-energy behavior of the electric tidal tensors $E_{\alpha\beta}$, $E_{\alpha\beta\gamma}$, $\dot E_{\alpha\beta}$ is, respectively, the same as the one of their  magnetic counterparts,
$B_{\alpha\beta}$, $B_{\alpha\beta\gamma}$, $\dot B_{\alpha\beta}$. This can be understood from the variance property of those tensors
 under boosts of $u_1$, when keeping $u_2$ fixed. For instance, 
 $K\equiv E_{\alpha\beta}(u_1)E^{\alpha\beta}(u_1) - B_{\alpha\beta}(u_1)B^{\alpha\beta}(u_1)$ is a 
 scalar\footnote{Actually it is proportional to the Kretschmann scalar.} invariant under boosts of $u_1$, i.e., independent of $u_1$. 
 This implies that,  contrary to the separate components of $E_{\alpha\beta}(u_1)$ and $B_{\alpha\beta}(u_1)$, $K$ is not amplified by a 
 factor $\gamma^2$, when boosting $u_1$ by a factor $\gamma$ with respect to $u_2$.
 It is instead independent of $\gamma$ and proportional to  $1/[R_2(z_1(\tau_1))]^6$. As a consequence, the  integral  
 $\langle K \rangle \sim \oint \frac{d\tau_1}{R_2^6(z_1(\tau_1))}\sim \frac{1}{p_\infty b^5}$. This  is also
linked to the fact that, as $\gamma \to \infty$, the Riemann curvature generated by $u_2$ and observed by $u_1$  
is of the null (wave-like) type~\cite{Aichelburg:1970dh}.
 
\section{Transcription of the tidal actions within the EOB  formalism}

In Refs. \cite{Damour:2009wj,Bini:2012gu} all consideration were limited to circular motions, and  tidal effects were translated into
a modification of the main radial potential $A=- g_{00}^{\rm eff}$ entering the general EOB mass-shell constraint \eqref{mass-shell}.
In the present, hyperbolic-motion, setting, it is more appropriate to translate tidal effects into an additional momentum-dependent
contribution to the $Q$ term in the general EOB mass-shell condition \eqref{mass-shell}. 
More precisely, we consider here a mass-shell condition of the form
\beq
\label{Q_tidal_first}
g^{\mu\nu}_{\rm Schw}P_\mu P_\nu +\mu^2 + Q_{\nu} + Q_{\rm tidal}=0\,.
\eeq
In Eq. \eqref{Q_tidal_first} we have used as EOB effective metric the Schwarzschild metric of mass $M$ (so that it incorporates the  full 1PM gravitational interaction \cite{Damour:2016gwp}); $P_\mu$ denotes $P^{\rm eob}_\mu$ and $Q=Q_{\nu} + Q_{\rm tidal}$  
is the sum of two types of contributions. The first one, $Q_{\nu}$,  represents
the post-1PM, and actually, post-Schwarzschild, effects due to PM gravity \cite{Damour2018},
\beq\label{QH}
{Q}_{\nu}(u, H_S)= u^2 Q_{\nu 2}(H_S)+  u^3 Q_{\nu 3}(H_S) + O(G^4),
\eeq
where $u\equiv GM/R_{\rm eob}$, and where $H_S$ denotes the Schwarzschild Hamiltonian
(say in Schwarzschild coordinates), i.e.
\beq
\label{sch_ham}
H_S(u,P_R,P_\varphi) =\sqrt{A(R) \left(  \frac{P_R^2}{B(R)}+ \frac{P_{\varphi}^2}{R^2} + \mu^2 \right)}\,,
\eeq
with $A(R)=1/B(R)=1-2u$.
The value of the 2PM, $O(G^2)$, term $ u^2 Q_{\nu 2}(H_S)$ has been computed in Ref. \cite{Damour2018},
while the value of the 3PM,  $O(G^3)$, term $ u^3 Q_{\nu 3}(H_S)$ is still a matter of debate \cite{Bern:2019nnu,Bern:2019crd,Antonelli:2019ytb,Damour:2019lcq}.
The additional, tidal-related contribution $Q_{\rm tidal}$ in Eq. \eqref{Q_tidal_first} is given by a sum of contributions corresponding to
all the different terms in the non-minimal worldline action \eqref{S_non_min_eq}.
When using the energy gauge of Ref. \cite{Damour2018}, each tidal term scaling like
the time integral of some power of the real interbody distance, say $\propto \int d\tau_A R_{12}^{- n}$ can be made to correspond
to an energy-dependent contribution $\propto u^n$ in $Q_{\rm tidal}$. It is convenient to work with the following dimensionless rescaled version of  $Q_{\rm tidal}$,
\beq
 {\widehat Q}_{\rm tidal} \equiv \frac{Q_{\rm tidal}}{\mu^2}\,.
\eeq
We will therefore be considering an energy dependent ${\widehat Q}_{\rm tidal}$ of the form 
\begin{eqnarray}
\label{q_defs_var}
{\widehat Q}_{\rm tidal}(u, \hat H_{\rm eff})&=&
- u^6  \left[ q^{E^2}_1(\hat H_{\rm eff})+ q^{B^2}_1(\hat H_{\rm eff})\right]\nonumber\\
&-&  u^8  \left[ q^{\tilde E^2}_1(\hat H_{\rm eff})+ q^{\tilde B^2}_1(\hat H_{\rm eff})\right]\nonumber\\
&-&  u^8  \left[ q^{\dot E^2}_1(\hat H_{\rm eff})
+  q^{\dot B^2}_1(\hat H_{\rm eff})\right]\nonumber\\
&-&  u^9  \left[ q^{E^3}_1(\hat H_{\rm eff})
+q^{ B^3}_1(\hat H_{\rm eff})\right]+\ldots\nonumber\\
& + & 1 \leftrightarrow 2\,. 
\end{eqnarray}
The precise choice of the energy argument $\hat H_{\rm eff}$ entering ${\widehat Q}_{\rm tidal}(u, \hat H_{\rm eff})$ 
is a matter of choice. One could take for $\hat H_{\rm eff}$ the conserved Hamiltonian associated with the $\nu$-deformed mass-shell condition $g^{\mu\nu}_{\rm Schw}P_\mu P_\nu +\mu^2 + Q_{\nu}=0$, or simply the Schwarzschild Hamiltonian $H_S$, Eq. \eqref{sch_ham},  associated with the $\nu$-undeformed mass-shell condition $g^{\mu\nu}_{\rm Schw}P_\mu P_\nu +\mu^2=0$. In the present paper, as we will work to leading-order PM accuracy, this choice will not matter and it will turn out that we can simply use for 
$\hat H_{\rm eff}$ the free-motion effective Hamiltonian $\hat H^{\rm eff}= \hat H^{\rm eff}_{\rm free} + O(G)$, with
$\hat H^{\rm eff}_{\rm free}= \mu^{-1}\sqrt{\mu^2+P_R^2+\frac{P_\phi^2}{R^2}}$.

The tool for converting real action terms into effective additional $Q$ terms is given by Eq. \eqref{eq_def_Q} above
that we rewrite here as
\beq
\label{eq_def_Q_new}
S^{\rm tidal}(E,P_\phi)=-\frac12 \int d\sigma_{(0)} Q^{\rm tidal}\,,
\eeq
where, using the free-motion effective Hamiltonian $H^{\rm eff}_0= \sqrt{\mu^2+P_R^2+\frac{P_\phi^2}{R^2}} + O(G)$, 
and $A=1-2u =1 +O(G)$, one finds [with a plus (minus) sign for the incoming (outgoing) radial integral]
\begin{eqnarray}
d\sigma_{(0)}&=&\frac{A dR^{\rm eob} }{  H^{\rm eff}_0 \frac{\partial H^{\rm eff}_0}{\partial P_{R}^{(0) {\rm eob}}  }}\nonumber\\
&=&\pm \frac{GM}{\mu}\frac{du}{u^2 \sqrt{(\gamma^2-1)-j^2u^2}} + O(G).
\end{eqnarray}
Here we introduced the dimensionless angular momentum
\beq
j \equiv \frac{P_\phi}{GM\mu}\,,
\eeq
and denoted the constant value  of the effective energy $E_{\rm eff}=H^{\rm eff}_0$ simply as $\mu \gamma$, i.e.,  $\gamma=\hat E_{\rm eff}$.
[Indeed, asymptotically the conserved effective EOB energy is equal to the Lorentz gamma factor, i.e. to 
$- (P_1 \cdot P_2)/(m_1 m_2)$.] Therefore, denoting $u_{\rm max}=\frac{\sqrt{\gamma^2-1}}{j}$, we find for the dimensionless integrated radial action
\beq
\label{eq_def_Q_new2}
\widehat S_R^{\rm tidal}\equiv \frac{S_R^{\rm tidal}}{GM \mu}=-  \int_{0}^{u_{\rm max}} \frac{ \widehat Q^{\rm tidal}(u) du}{u^2 \sqrt{(\gamma^2-1)-j^2u^2}} \,.
\eeq
When $\widehat Q^{\rm tidal}(u,\gamma)=-q_n(\gamma) u^n$ (where the argument $\hat H_S$ entering Eq. \eqref{q_defs_var} has been replaced by $\gamma$) we find (as long as $n>1$)
\beq
\label{eq_def_Q_new3}
\widehat S_R^{\rm tidal}=+I_n \, q_n(\gamma)  \frac{(\gamma^2-1)^{\frac{n-2}{2}}}{j^{n-1}} \,,
\eeq
where
\beq
I_n=\int_0^1 \frac{x^{n-2}}{\sqrt{1-x^2}}dx= \frac{\sqrt{\pi}}{2}\frac{\Gamma\left(\frac{n-1}{2} \right)}{\Gamma\left(\frac{n}{2} \right)}\,.
\eeq

Let us focus on the dominant tidal contribution associated with the electric quadrupole tensor.
In that case, $n=6$ and the left-hand-side of Eq. \eqref{eq_def_Q_new3} is
\beq
\frac14 \frac{m_2}{m_1} \hat \mu_1^{(2)}\, \left(\frac{GM}{b} \right)^5\, {\mathcal F}_{E^2} (\gamma)\,, 
\eeq
where 
\beq
{\mathcal F}_{E^2}(\gamma) =\frac{9 \pi(35\gamma^4-30\gamma^2+11)}{64\sqrt{\gamma^2-1}} \,,
\eeq
and where we introduced the dimensionless version of $\mu_1^{(2)}$, etc. defined as follows
\begin{eqnarray}
\hat\mu_1^{(2)}&=&\frac{G\mu_1^{(2)}}{(GM/c^2)^5}\,,\nonumber\\
\hat\mu_1^{(2)}{}'&=&\frac{G \mu_1^{(2)}{}'}{(GM/c^2)^7}\,,\nonumber\\
\hat\mu_1^{(3)}&=&\frac{G \mu_1^{(3)}}{(GM/c^2)^7}\,,
\end{eqnarray} 
etc.

On the other hand, the right-hand-side is, using $I_6=\frac{3\pi}{16}$, 
\beq
\frac{3\pi}{16} q_1^{E^2}  \frac{(\gamma^2-1)^{2}}{j^5}\,.
\eeq
Using the link between $b$ and $j$ \cite{Damour2018}
\beq
\frac{GM}{b}=\frac{\sqrt{\gamma^2-1}}{h(\gamma, \nu) j}\,,
\eeq
where (using $\hat H_{\rm eff}=\gamma$)
\beq
h(\gamma, \nu)=\frac{E_{\rm real}}{M}=\sqrt{1+2\nu(\gamma-1)}\,,
\eeq
one obtains
\beq
q_1^{E^2}=3\frac{m_2}{m_1} \hat \mu_1^{(2)} \frac{35\gamma^4-30\gamma^2+11}{16 h^5(\gamma, \nu)}   \,.
\eeq
This is the contribution to the tidal influence of body 2 on body 1.
Therefore, the coefficient of $-u^6$ in $\hat Q_{\rm tidal}$ due to the quadrupolar-electric 
interaction between the two bodies will be the $1\rightarrow 2$ completion of this result, namely
\begin{eqnarray}
\label{eq:110_bis}
q_{{\rm T}}^{E^2}&=& q_1^{E^2}+q_2^{E^2}\nonumber\\
&=& 3\left(\frac{m_2}{m_1} \hat \mu_1^{(2)}+\frac{m_1}{m_2} \hat \mu_2^{(2)}\right) \frac{35\gamma^4-30\gamma^2+11}{16 h^5(\gamma, \nu)} \nonumber\\
&\equiv & 
3\hat \mu_*^{(2)}\frac{35\gamma^4-30\gamma^2+11}{16 h^5(\gamma, \nu)}
\,,
\end{eqnarray}
where we defined
\beq
\hat \mu^{(2)}_* \equiv \frac{m_2}{m_1}\hat \mu _1^{(2)}+\frac{m_1}{m_2}\hat \mu _2^{(2)}\,.
\eeq
To complete our EOB reformulation of tidal effects within the PM framework let us connect it to the previous PN-based EOB formulation.
The latter \cite{Damour:2009wj} was focusing on circular motions and was describing tidal effects by means of an additional radial function $A_{\rm tidal}(u)$ in the main EOB radial potential.
This is equivalent to describing tidal effects by   the following   squared  effective Hamiltonian 
\beq
H_{\rm eff}^2 = (A(R)+A_{\rm tidal}(R))\left[\mu^2 +\frac{P_R^2}{B}+\frac{P_\phi^2}{R^2} + Q_\nu\right]\,.
\eeq
By contrast our present approach consists of describing tidal effects by  
\beq
H_{\rm eff}^2 =  A(R) \left[\mu^2+\frac{P_R^2}{B}+\frac{P_\phi^2}{R^2}+ Q_\nu + Q_{\rm tidal}  \right]\,.
\eeq
Comparing the two approaches we see that our $Q_{\rm tidal} $ contribution can be translated (along circular orbits) in the following  equivalent tidal potential 
\beq
A_{\rm tidal}(u)=\left[\frac{A(u)}{1+\frac{p_r^2}{B}+ p_\phi^2 u^2+\widehat Q_\nu} \widehat Q_{\rm tidal} \right]^{\rm circ}\,,
\eeq
where now both $Q_{\rm tidal}$ and the phase space variables have been rescaled, notably, $p_r=P_R/(\mu)$, $p_\phi=P_\phi/(GM\mu)=j$.

Our PM approach to tidal effects, when restricting to electric quadrupolar effects,  describes them by 
the $H_{\rm eff}$-dependent contribution 
\beq
\hat Q^{\rm tidal}(\hat H_{\rm eff})=-  q_{\rm T}^{E^2}(\hat H_{\rm eff}) u^6\,,
\eeq
where $q_{\rm T}^{E^2}(\gamma)$ is given by Eq. \eqref{eq:110_bis}.
Interpreting $\hat H_{\rm eff}^2$  as denoting $\hat H_{\rm eff}^2=A(1+1+\frac{p_r^2}{B}+ p_\phi^2 u^2+\widehat Q_\nu)$, this result is equivalent to an effective energy-dependent $A_{\rm tidal}$ potential equal to
\beq \label{Atidal}
A_{\rm tidal}(u,\hat H_{\rm eff}) =-u^6\frac{A^2(u) q_{\rm T}^{E^2}(\hat H_{\rm eff})}{\hat H_{\rm eff}^2}\,.
\eeq
One cannot directly compare the full energy-dependence predicted by this lowest-PM accuracy result to previous PN-based results
which were  PN-corrected, i.e., which included  combinations of both $p_\infty^2=\gamma^2-1$ and $u=GM/R$ as corrections up to the
2PN level \cite{Bini:2012gu}.
One would need to compute higher PM gravity corrections (i.e. fractional corrections to Eq. \eqref{eq:110_bis}
involving powers of $u=GM/R$) to our PM-based result for doing a meaningful comparison. 
However, we will compare our new
formulation to the previous one, at the {\it Newtonian} level, in the next section.

\section{Tidal contribution to the scattering angle and to periastron precession}

As we have seen above, the leading PM-order contribution to the tidal action  is given by
\begin{eqnarray}
\frac{S_R^{E^2}(E,P_\phi)}{GM\mu}&=& \frac14 \hat \mu^{(2)}_*{\mathcal F}_{E^2}(\gamma) \left(\frac{GM}{b} \right)^5\nonumber\\
&=&  \frac14 \hat \mu^{(2)}_*{\mathcal F}_{E^2}(\gamma)\frac{p_\infty^5}{h^5 j^5}\nonumber\\
&=& \frac{9\pi }{16} \hat \mu^{(2)}_*  \left(1+ \frac{5}{2}p_\infty^2+\frac{35}{16}p_\infty^4\right)\, \frac{p_\infty^4}{h^5 j^5}
\,. \nonumber\\
\end{eqnarray}
Using Eq. \eqref{chi_1_ref} gives the corresponding leading PM-order tidal contribution to the scattering angle $\chi$ as a function of $\gamma$ and $j$, where we have used  $\gamma=\hat H_{\rm eff}$ and $j=P_\phi/(GM\mu)$
\begin{eqnarray}
\label{chi_E_2_expl}
\chi_{E^2}(p_\infty,j) &=& \frac54 \hat \mu^{(2)}_* \frac{ {\mathcal F}_{E^2}(\gamma) p_\infty^5}{h^5 j^6}\nonumber\\
&=& \frac{45 \pi}{256} \hat \mu^{(2)}_*   (16+ 40p_\infty^2+35p_\infty^4)  \frac{p_\infty^4}{h^5 j^6}\,. \nonumber \\
\end{eqnarray}
Let us recall the beginning of the PM expansion of the scattering angle due to Einstein gravity
\begin{eqnarray} \label{chiPMbis}
\frac12 \chi^{\rm }(\gamma, j;\nu) &=& \frac{\chi_{1}(\gamma, \nu)}{j} + \frac{ \chi_{2}(\gamma, \nu)}{j^2} 
+  \frac{\chi_{3}(\gamma, \nu)}{j^3} \nonumber\\ 
&&+  \frac{\chi_{4}(\gamma, \nu)}{j^4}  + \cdots,
\end{eqnarray}
where
\beq \label{chi1pm}
\chi_{1}(\gamma,\nu)=   \frac{2 \gamma^2-1}{\sqrt{\gamma^2-1}} \,,
\eeq
does not depend on $\nu$ and
\beq \label{chi2pm}
\chi_{2}(\gamma, \nu)=   \frac{3 \pi}{8} \frac{(5 \, \gamma^2-1)}{h(\gamma,\nu)}\,.
\eeq
In the HE limit $\gamma \to \infty$ this 2PM-accurate scattering angle has
a finite limit (which is independent of the symmetric mass ratio $\nu$) if the ratio 
\beq
\alpha=\frac{\gamma}{j}\,,
\eeq
is kept fixed (and small).
Namely,
\begin{eqnarray} 
\label{chiPMbis-n}
\frac12 \chi^{\rm }(\gamma, j; \nu) &\overset{\rm HE}{=}& 2 \alpha + \frac{15\pi}{8}\frac{\alpha^2}{h}+O(G^3)\,,
\end{eqnarray}
where the term of order $\alpha^2$ is negligible because $h \approx \sqrt{2\nu\gamma}\to \infty$.
There is a current debate about the recently computed 3PM $O(G^3)$ contribution \cite{Bern:2019nnu,Bern:2019crd} which yields a 
divergent $G^3$ contribution, $O(\alpha^3\ln(\gamma) \to \infty$ as  $\gamma\to \infty$, while the computation of Ref. \cite{Amati:1990xe} (see also \cite{DiVecchia:2019kta}) got a finite $O(\alpha^3)$ correction in the massless limit. [The massless limit $m_1, m_2 \to 0$ corresponds to $\gamma = - (P_1 \cdot P_2)/(m_1 m_2) \to \infty$.]
 See also the conjectured 3PM dynamics of Ref. \cite{Damour:2019lcq} leading to a finite $O(\alpha^3)$. In addition, both Refs. \cite{Amati:1990xe} and \cite{Damour:2019lcq} 
argued 
that the HE limit of $\chi/2$ should be of the form $2 \alpha +c_3\alpha^3 +c_5 \alpha^5+\ldots$, i.e., with odd powers only.

Let us then consider the high-energy limit of $\chi^{E^2}$ when $\alpha$ is kept fixed.
It reads
\beq  \label{chi_to_complete}
\chi^{E^2}  \overset{\rm HE}{=}
\frac{1575}{256} \pi  \hat \mu_*^{(2)}  \frac{\gamma^2}{h^5}\alpha^6\,.
\eeq
It contains a factor $\alpha^6$ as expected from a $\sim j^{-6}$ effect. We note, however, the following limiting property of the energy-dependent ratio multiplying $\alpha^6$
\beq
\frac{\gamma^2}{h^5}=\frac{\gamma^2}{[1+2\nu(\gamma-1)]^{5/2}} \overset{\rm HE}{=} \frac{1}{(2\nu)^{5/2}}\frac{1}{\gamma^{1/2}}\to 0\,.
\eeq
From this point of view, we conclude that, in the HE limit, the tidal corrections to scattering behave in the way expected, i.e.
similarly to  the non-tidal $\alpha^6$-correction coming from Einstein gravity. Indeed, one expects the latter to be suppressed compared 
to the terms involving odd powers of $\alpha$ (in a way similar to the $O(\alpha^2)$ contribution discussed above), so as to
leave a final result independent of the masses (and of the internal structures) of the scattered HE bodies.

Let us  note in passing that the HE behavior of tidal effects is also potentially important when considering circular motions.
Indeed, Ref. \cite{Bini:2012gu} has pointed out the possible existence of a power-law divergence as the circular motion is formally  allowed to approach the lightring, i.e., a circular orbit where $\hat H_{\rm eff}^2=\gamma^2$ goes to infinity. 
Even if such an orbit is never physically reached during the inspiral of a real binary system,  such a power-law blow-up would imply an increase of the strength of tidal effects during the last inspiraling orbits before merger which seems to be needed to get a good agreement with numerical simulations. [For more discussions of this issue see Refs. \cite{Bernuzzi:2014owa,Steinhoff:2016rfi}.] However, as already
mentioned, our current 1PM-accurate result would need to be improved by $\sim GM/R + (GM/R)^2 + \ldots$ PM corrections to
meaningfully discuss the HE behavior happening near the lightring.

Recently, Ref. \cite{Kalin:2019rwq} has pointed out a simple link between scattering angle and periastron precession, namely
\beq \label{Porto}
\Delta \Phi(E,P_\phi)=\left[ \chi(E,P_\phi)+\chi(E,-P_\phi) \right]^{\rm analytically \, continued}\,,
\eeq
under the assumption that one can define a suitable analytic continuation of the energy $E$  from the scattering domain, $E \geq M$,   to the bound-state  one, $E \leq M$. In terms of $\gamma= E_{\rm eff}/\mu$ this corresponds to a continuation
from $\gamma \geq1 $ to $\gamma\leq 1$, while in terms of the more directly relevant variable $p_\infty = \sqrt{\gamma^2-1}$,
this corresponds to a {\it Wick rotation}  of $p_\infty$ from the real axis to the imaginary one.
If one  applies this prescription to the above lowest-order PM estimate of the tidal scattering angle $\chi^{E^2}(p_\infty,j)$ one 
{\it formally} gets a corresponding periastron precession equal to
\beq
\label{1PMperiprec}
\Delta \Phi_{E^2}^{ 1 \rm PM}(p_\infty,j)=
 \frac{45 \pi}{128} \hat \mu^{(2)}_*   (16+ 40p_\infty^2+35p_\infty^4)  \frac{p_\infty^4}{h^5 j^6}\,,
\eeq
without encountering ambiguities in the analytic continuation in $p_\infty$ because the odd power of $p_\infty$ contained in ${\mathcal F}_{E^2}$ has been cancelled by the $p_\infty^5$ factor.

However, one cannot directly compare the full structure of the formal 1PM expression \eqref{1PMperiprec} to any known, well-defined 
periastron advance result. Indeed, Eq. \eqref{1PMperiprec} is the first term in an expansion in 
powers of $1/j \propto G$, while keeping
 fixed $p_\infty$. The missing fractional corrections to \eqref{1PMperiprec} include, in particular, powers of 
 \beq \label{epsilon}
\epsilon\equiv \frac{1}{p_\infty j}\,.
\eeq
[The notation $\epsilon$ introduced here should not be confused with the use
of $\epsilon$ as a generic small parameter in Sec. II above.]
In other words, the 1PM expression \eqref{1PMperiprec} makes sense only if $\epsilon \ll1$.
On the other hand, the periastron advance $\Delta \Phi_{E^2}$ is an observable which makes sense only
for ellipticlike, bound orbits, i.e. in the case where the Newtonian eccentricity, $e$, whose square can be defined as (see below)
\beq \label{e2}
e^2\equiv 1+p_\infty^2 j^2 = 1 + \frac1{\epsilon^2} \,,
\eeq
is {\it smaller} than 1. This conflicts with the domain of validity $\epsilon \ll1$, of the PM expansion, which implies $e \gg1$.

However, if we restrict ourselves to the PN regime, $c \to \infty$, in which $ p_\infty  \propto \frac1c \ll1$, while $ j \propto c \gg1$,
keeping fixed the product $ p_\infty j$ (and therefore the eccentricity), we can {\it interpolate} between the PM domain of
validity, $e \gg1$, and the elliptic-motion domain, $e <1$, where $\Delta \Phi_{E^2}$ is defined. Let us then consider 
 the non-relativistic (Newtonian-level) limits (obtained using $p_\infty \ll 1$ and $h \approx 1$) 
of our above 1PM results for $\chi$, and its formal counterpart $\Delta \Phi$ (Eq. \eqref{1PMperiprec}),  namely
\beq
\label{N1PMchi}
\chi_{E^2}^{ 1 \rm PM \cap Newton}(p_\infty,j)=
 \frac{45 \pi}{16} \hat \mu^{(2)}_*     \frac{p_\infty^4}{ j^6}\,,
\eeq
and
\beq
\label{N1PMperiprec}
\Delta \Phi_{E^2}^{ 1 \rm PM \cap Newton}(p_\infty,j)=
 \frac{45 \pi}{8} \hat \mu^{(2)}_*     \frac{p_\infty^4}{ j^6}\,.
\eeq
We are going to compare these expressions to the corresponding complete Newtonian-level predictions for the (quadrupolar) tidal 
contributions, $\chi_{E^2}^{\rm Newton}$ and
$\Delta \Phi_{E^2}^{\rm Newton}$, to the scattering angle and to periastron precession.

As far as we know, while $\Delta \Phi_{E^2}^{\rm Newton}(p_\infty,j)$ is known from classic works in Newtonian gravity \cite{sterne}, 
the corresponding scattering angle $\chi_{E^2}^{\rm Newton}(p_\infty,j)$  has never been obtained in the literature. We are
going to derive $\chi_{E^2}^{\rm Newton}(p_\infty,j)$, and the corresponding $\Delta \Phi_{E^2}^{\rm Newton}(p_\infty,j)$,
and then compare these Newtonian-level results to the above $1 \rm PM \cap Newton$ expressions \eqref{N1PMchi}, \eqref{N1PMperiprec}.

To derive $\chi_{E^2}^{\rm Newton}(p_\infty,j)$ (and the corresponding $\Delta \Phi_{E^2}^{\rm Newton}(p_\infty,j)$)
we must consider the effect of adding a $O(1/R^6)$ perturbation to the Newtonian potential.
We can easily do that within the EOB framework by considering a  squared effective EOB Hamiltonian of the form
\beq \label{HfNtidal}
 (\hat H_{\rm eff}^2)^{\rm Newton}=1+ p_r^2 +j^2u^2-2u - C u^6\,.
\eeq
Such an effective EOB Hamiltonian is obtained from Eq. \eqref{Q_tidal_first} by treating the Schwarzschild piece of the
mass-shell contraint to leading Newtonian order (i.e. using $A=1-2u$ and $B=1$), by neglecting the 2PM correction $Q_{\nu}$,
and by adding only the leading-order $O(u^6)$ tidal term from $Q_{\rm tidal}$. From Eqs. \eqref{Atidal} and \eqref{eq:110_bis}, 
the coefficient $C$ in front of $- u^6$ is the Newtonian limit of $ q_{\rm T}^{E^2}(\hat H_{\rm eff})$, namely
\beq
C= 3 \, \hat \mu_*^{(2)}\,.
\eeq
The conserved energy of $ (\hat H_{\rm eff}^2)^{\rm Newton}-1 $ is $\gamma^2- 1 = p_\infty^2$.

We can then apply the same approach as above to the effective Hamiltonian \eqref{HfNtidal}.
Instead of Eq. \eqref{eq_def_Q_new2}, we now get the Newtonian-level tidal action 
\beq
\label{SNtidal}
\widehat S_{\rm Newton}^{\rm tidal} =-  \int_{0}^{u_{+}} \frac{ \widehat Q^{\rm tidal}(u) du}{u^2 \sqrt{p_\infty^2-j^2u^2 + 2u}} \,.
\eeq
where  $\widehat Q^{\rm tidal}(u) = - C u^6$, and where $u_{+}$ is the positive root of the squared denominator.

The computation of the integral \eqref{SNtidal} is elementary. In terms of the above-defined $\epsilon$, Eq. \eqref{epsilon},
it can be rewritten as
\beq \label{SNtidal2}
\widehat S_{\rm Newton}^{\rm tidal} = C \frac{p_\infty^4}{j^5} I_4(\epsilon)\,,
\eeq
where
\beq
 I_4(\epsilon)= +\int_{0}^{x_{+}(\epsilon)} \frac{ x^4 dx}{\sqrt{1-x^2 + 2 \epsilon x}} \,,
\eeq
with $x_{+}(\epsilon)= \epsilon + \sqrt{1+\epsilon^2}$.
The explicit value of $ I_4(\epsilon)$ is
\beq \label{I4}
I_4(\epsilon)=\left(\frac{3}{8}+\frac{15}{4}\epsilon^2+\frac{35}{8}\epsilon^4\right) B(\epsilon)
+\frac{55}{24}\epsilon+\frac{35}{8}\epsilon^3\,,
\eeq
with $B(\epsilon) \equiv {\rm arctan}(\epsilon) +\frac{\pi}{2}$.

From the action \eqref{SNtidal2}, one computes $\chi$ by differentiating with respect to $j$:
\beq
\chi_{E^2}^{\rm Newton}(p_\infty,j) = - \frac{\partial}{\partial j} \widehat S_{\rm Newton}^{\rm tidal}(p_\infty,j) \,.
\eeq
This leads to the following result for $\chi$,
\beq \label{chiNfull}
\chi_{E^2}^{\rm Newton}(p_\infty,j)= C \frac{p_\infty^4}{j^6}\left[ 5 I_4(\epsilon)+ \epsilon \frac{\partial  I_4(\epsilon)}{\partial \epsilon}\right]\,,
\eeq
whose explicit form reads
\begin{eqnarray} \label{chiNexpl}
\chi_{E^2}^{\rm Newton}(p_\infty,j)&=& C \frac{p_\infty^4}{j^6}\left[
\left(\frac{15}{8}+\frac{105}{4}\epsilon^2+\frac{315}{8}\epsilon^4\right) B(\epsilon)\right.\nonumber\\
&&\left.+\frac{\epsilon (113+420\epsilon^2+315\epsilon^4)}{8(1+\epsilon^2)}  
\right]\,.
\end{eqnarray}
This expression is a complicated function of $\epsilon$. At this stage, one must remark that $\epsilon= G m_1 m_2/(p_\infty P_\phi)$
is of order $O(G)$. Our previous 1PM result was obtained at leading order in the expansion in powers of $G$. This corresponds to taking
the leading order in the expansion of $\chi_{E^2}^{\rm Newton}$ in powers of $\epsilon$. Alternatively (and more physically),
we can remark that the quantity $e$, defined  above by Eq. \eqref{e2},
is the Newtonian eccentricity of the orbit, corresponding to the existence of two roots of $p_r^2$ (corresponding to the squared
denominator of Eq. \eqref{SNtidal}). More precisely, $e^2$  is larger than 1 when $p_\infty$ is real, so that $p_\infty^2>0$ (hyperbolic
motion), and smaller than 1 when $p_\infty$ is purely imaginary, so that $p_\infty^2<0$ (elliptic motion).
The leading order PM approximation (small scattering angle) corresponds to $e \to \infty$, which indeed corresponds to $\epsilon \to 0$.

In the $\epsilon \to 0$ limit, it is enough to use the first term in the $\epsilon$ expansion of $I_4(\epsilon)$:
\beq
I_4(\epsilon)=\frac{3}{16}\pi+\frac{8}{3}\epsilon+\frac{15}{8}\pi\epsilon^2+8\epsilon^3+\frac{35}{16}\pi\epsilon^4+O(\epsilon^5)\,,
\eeq
namely $I_4(0)=\frac{3}{16}\pi$. Inserting this value in Eq. \eqref{chiNfull} (or taking the $\epsilon \to 0$ limit of \eqref{chiNexpl}) yields
\beq
\chi_{E^2}^{{\rm Newton \, large} \,e}=\frac{15 \pi}{16} C \frac{p_\infty^4}{j^6}= \frac{45 \pi}{16} \hat \mu_*^{(2)} \frac{p_\infty^4}{j^6}\,,
\eeq
in agreement with Eq. \eqref{N1PMchi}.

Let us now consider the periastron advance $\Delta \Phi$, as obtained from Eq. \eqref{Porto}.
 The analytic continuation from hyperbolic to elliptic motions 
involves Wick rotating $p_\infty$, and therefore $\epsilon=1/(p_\infty j)$, from the real axis to the imaginary axis.
In addition, Eq. \eqref{Porto} involves taking (twice) the even part with respect to $j$, i.e., taking (twice) 
the even part with respect to $\epsilon$. Finally, this leads to an expression for  $\Delta \Phi$ obtained
from Eq. \eqref{chiNfull} by replacing $I_4(\epsilon)$ by
\beq
 J_4(\epsilon)= I_4(\epsilon)+ I_4(-\epsilon) \, . 
 \eeq
 This replacement drastically simplifies $I_4(\epsilon)$ into
 \beq
  J_4(\epsilon) =\pi \left(\frac{3}{8}+\frac{15}{4}\epsilon^2+\frac{35}{8}\epsilon^4\right) \,.
 \eeq
 This then yields
 \begin{eqnarray} \label{PhiNfull}
\Delta \Phi_{E^2}^{\rm Newton}(p_\infty,j) &=& C \frac{p_\infty^4}{j^6}\left[ 5 J_4(\epsilon)+ \epsilon \frac{\partial  J_4(\epsilon)}{\partial \epsilon}\right] \nonumber\\
&=& \frac{15 \pi}{8} C \frac{p_\infty^4}{j^6} \left[1+ 14 \epsilon^2 + 21 \epsilon^4 \right]\,, \nonumber\\
&=& 15 \pi C \frac{p_\infty^4}{j^6 (1-e^2)^2} \left[1+ \frac32 e^2 + \frac18 e^4 \right]\,.\nonumber
\end{eqnarray}
The analytic continuation to imaginary values of $p_\infty$ is then unambiguous because the above expression is a function
of $p_\infty^2$.

Finally, using $p_\infty^2 j^2= e^2-1$, and $C=  3 \, \hat \mu_*^{(2)}$, one gets
\beq \label{PhiN}
\Delta \Phi_{E^2}^{\rm Newton}(p_\infty,j) = \frac{45  \pi  \hat \mu_*^{(2)}}{j^{10}} \left[1+ \frac32 e^2 + \frac18 e^4 \right] \,.
\eeq
On the one hand, this expression agrees with the classic Newtonian result of Ref. \cite{sterne}, when using the link \cite{Damour:2009vw}
\beq
G \mu^{(2)}_A=  \frac{2 }{3} k^{(2)}_A  R_A^5\,.
\eeq
On the other hand, if we {\it formally} consider the {\it large eccentricity limit}, $e \gg1$, of the expression \eqref{PhiN}
(though it is physically defined only when $e <1$), one gets
\begin{eqnarray}
\Delta \Phi_{E^2}^{\rm Newton}(p_\infty,j) &&\overset{ e \gg 1}{=}  \frac{45  \pi  }{ 8 } \hat \mu_*^{(2)} \frac{e^4}{j^{10}} \nonumber\\
&&\overset{ e \gg 1}{=}  \frac{45 \pi}{8} \hat \mu^{(2)}_*     \frac{p_\infty^4}{ j^6}\,.
\end{eqnarray}
On the second line, we have used the definition \eqref{e2} of $e^2$, leading to $e^4 \approx p_\infty^4 j^4$ in the large-eccentricity limit.
We  see that the latter final expression agrees with the formal $ 1 \rm PM \cap Newton$ expression \eqref{N1PMperiprec}.

This exercize has highlighted the fact that, in spite of the simple formal link \eqref{Porto},
there is a long theoretical distance separating the PM-expansion of the scattering angle (valid in the
large eccentricity limit, $e \gg 1$), and the PN-expansion of the periastron advance (meaningful only for $e <1$).

\section{Concluding remarks}

We have extended the post-Minkowskian approach to the computation of tidally interacting binary systems.
Our computation used the effective field theory description of 
tidally interacting bodies, and was simplified by using general properties of perturbed actions. 
We computed several tidal invariants (notably the integrated quadrupolar
and octupolar actions) at the first post-Minkowskian order, and derived the corresponding contributions 
to the scattering angle, and to the periastron advance. We showed also how to transcribe our post-Minkowskian tidal
results in the effective one body formalism, using the same type of energy gauge that was recently used in the 
post-Minkowskian approach to the dynamics of point masses. It would be interesting to extend our computation
to higher post-Minkowskian levels so as, notably, to clarify the high-energy behavior of the tidal interaction of two bodies.

\section*{Acknowledgments}
D.B. thanks the International Center for  
Relativistic Astrophysics Network (ICRANet)  for  
partial support, and the Institut des Hautes Etudes 
Scientifiques (IHES) for warm hospitality
during the completion of the present project.

\appendix

\section{Four velocities, momenta and center-of-mass at the Minkowskian level}

\begin{figure}[h] 
\includegraphics[scale=0.70]{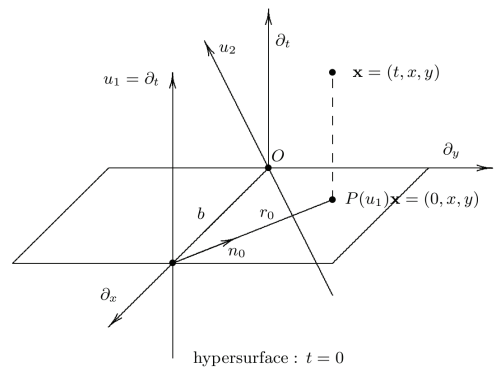}
\caption{\label{fig:1} The figure shows the choice of the the axes (adapted to $u_1^\mu$,
and to the vectorial impact parameter $b^\mu$), and the origin (located on ${\mathcal L}_2$),
  of the coordinates we use in the text.
 The $z$ coordinate of any point is assumed to be 0 and is omitted.}  
\end{figure} 

In the Minkowskian discussion of the two-body problem the following four-vectors are relevant
\begin{eqnarray}
u_1 &=& \partial_t\,, \nonumber\\
u_2 &=& \gamma \partial_t -\sqrt{\gamma^2-1}\partial_y\,, \nonumber\\
U&=&\frac{m_1 u_1 +m_2 u_2}{E_{\rm real}}
=\frac{X_1}{h}u_1+\frac{X_2}{h}u_2\,, 
\end{eqnarray}
where $u_1$ is the four-velocity of  body 1, $u_2$ that of  body 2,  $U$ that of the c.m. frame, and
\begin{eqnarray}
 E_{\rm real}&=&\sqrt{m_1^2+m_2^2+2m_1 m_2 \gamma}=Mh(\gamma,\nu)\,.
\end{eqnarray}
Here we used coordinates adapted to ${\mathcal L}_1$.

One can complete the unit timelike vectors $u_1$, $u_2$ and $U$ by corresponding  spatial, orthonormal vectorial frames 
(respectively orthogonal to $u_1$, $u_2$ and $U$)   as follows
\begin{enumerate}
  \item  Spatial, orthonormal frame completing $u_1$
\beq
e(u_1)_1=\partial_x\,,\qquad e(u_1)_2=\partial_y\,,\qquad e(u_1)_3=\partial_z\,.
\eeq
  \item Spatial, orthonormal frame  completing $u_2$
\begin{eqnarray}
\label{u_2_spat_frame}
e(u_2)_1&=&\partial_x\,,\nonumber\\ 
e(u_2)_2&=&-\sqrt{\gamma^2-1}\partial_t +\gamma \partial_y\,,\nonumber\\
e(u_2)_3&=&\partial_z\,.
\end{eqnarray} 
  \item Spatial, orthonormal frame  completing $U$
\begin{eqnarray}
\label{U_spat_frame}
e(U)_1&=&\partial_x\,,\nonumber\\
e(U)_2&=&-\sinh \alpha\partial_t +\cosh \alpha \partial_y\,,\nonumber\\
e(U)_3&=&\partial_z\,,
\end{eqnarray}
where
\begin{eqnarray}
\sinh \alpha &=& \frac{m_2 \sqrt{\gamma^2-1}}{E_{\rm real}}\,,\quad 
\cosh \alpha = \frac{m_1+m_2 \gamma}{E_{\rm real}}\,.
\end{eqnarray}
\end{enumerate}
Note the expressions of $\sinh\alpha$ and $\cosh \alpha$ in terms of $j$ and of the impact parameter $b$:
\beq
\sinh \alpha = \frac{Gm_2 j}{b}\,,\qquad \cosh\alpha =\frac{Gj}{b} \frac{m_1+m_2\gamma}{\sqrt{\gamma^2-1}}\,,
\eeq
implying
\beq
\frac{Gm_1  j}{b}=\sqrt{\gamma^2-1}\cosh\alpha -\gamma \sinh \alpha\,.
\eeq
With this notation  the center-of-mass 4-velocity $U$ reads
\beq
\label{U_def_in_coords}
U=\cosh \alpha \,\partial_t -\sinh \alpha \,\partial_y\,.
\eeq
These frames are obtained by boosting the spatial frame of $u_1$ into the local rest spaces of $u_2$ and $U$.
The spatial frame associated with $U$ has the peculiarity that one leg of the triad [$e(U)_2$] is aligned with the direction of the spatial momentum of each of the particles.

The spacetime vectorial frame  $(U, e(U)_1,e(U)_2,e(U)_3)$ is the c.m. frame. 
Decomposing $P_1=m_1 u_1$ and $P_2=m_2u_2$ along this frame gives
\begin{eqnarray}
P_1&=& m_1 u_1 =E_1 U+P \, e(U)_2\,,\nonumber\\
P_2&=& m_2 u_2 =E_2 U-P \, e(U)_2\,,
\end{eqnarray}
with $E_1+E_2= E_{\rm real}$, and
\begin{eqnarray}
E_1&=&\sqrt{m_1^2+P^2}= m_1 \cosh \alpha =m_1 \frac{m_2\gamma +m_1}{E_{\rm real}}\,,\nonumber\\
E_2&=&\sqrt{m_2^2+P^2}= m_2 \cosh \alpha'\equiv m_2 \frac{m_1\gamma +m_2}{E_{\rm real}}\,,\nonumber\\
P &=&\frac{m_1 m_2\sqrt{\gamma^2-1}}{E_{\rm real}}=m_1 \sinh \alpha =m_2 \sinh \alpha'\,. \nonumber\\
\end{eqnarray}
The Mandelstam variable $s$ associated with $P_1$ and $P_2$ reads
\beq
s=-(P_1+P_2)^2=E_{\rm real}^2\,.
\eeq

We also recall the following definitions for the spatial four-velocity\footnote{We denote some spacelike four-vectors by a boldface.} 
of each body seen in the rest frame of the other one, 
\begin{eqnarray}
{\mathbf u}_{21}&=&P(u_1)u_2=-\sqrt{\gamma^2-1}\, \partial_y\,,\nonumber\\ 
{\mathbf u}_{12}&=&P(u_2)u_1=\sqrt{\gamma^2-1} \, e(u_2)_2\,,
\end{eqnarray}
where we recall that $P(u)= I + u \otimes u$ denotes the projector orthogonal to the unit timelike vector $u$.
Here $e(u_2)_2$, defined in Eq. \eqref{u_2_spat_frame}, is the boosted $y$ axis in the local rest space of $u_2$ 
with $u_2\cdot e(u_2)_2=0$.
Therefore, 
\beq
{\mathbf u}_{12} =-(\gamma^2-1)\partial_t +\gamma \sqrt{\gamma^2-1}\partial_y\,,
\eeq
and
\beq
u_1\cdot {\mathbf u}_{12}=u_2 \cdot {\mathbf u}_{21}=\gamma^2-1\,.
\eeq
Note that  $|{\mathbf u}_{12}|= |{\mathbf u}_{21}| = \sqrt{\gamma^2-1}$ is equal to the EOB asymptotic momentum $p_\infty$.

For completeness, let us write the parametric equations of the two worldlines in the coordinate system
associated with ${\mathcal L}_1$
\begin{eqnarray}
z_1(\tau_1)&=& \tau_1 \partial_t +b\partial_x \,,\nonumber\\
z_2(\tau_2) &=& \gamma \tau_2 \partial_t -\sqrt{\gamma^2-1}\tau_2 \partial_y \,. 
\end{eqnarray}
We can also define  coordinates,
$(t_{\rm cm}, x_{\rm cm}, y_{\rm cm}, z_{\rm cm})$, adapted to the c.m. frame. They are related
to the coordinates $(t,x,y,z)$ adapted to ${\mathcal L}_1$ via
\begin{eqnarray}
t_{\rm cm}&=&\cosh \alpha \, t +\sinh \alpha \, y \,,\nonumber\\
x_{\rm cm}&=& x - \frac{E_1}{E_{\rm real}} b\,,\nonumber\\
y_{\rm cm}&=&\sinh \alpha \, t +\cosh \alpha \, y  \,,\nonumber\\
z_{\rm cm}&=& z\,,
\end{eqnarray}
 with inverse
\begin{eqnarray}
t&=&\cosh \alpha \, t_{\rm cm} -\sinh \alpha \, y_{\rm cm} \,,\nonumber\\
x&=& x_{\rm cm} + \frac{E_1}{E_{\rm real}} b\,,\nonumber\\
y&=&-\sinh \alpha \, t_{\rm cm} +\cosh \alpha \, y_{\rm cm}\, \nonumber\\
z&=& z_{\rm cm}\,,
\end{eqnarray}
As is standard, the origin of these coordinates has been taken as the center of energy of $z_1$ and $z_2$,
when viewed in the c.m. frame, and at the same c.m. time $t_{\rm cm}$.

Finally, the Pauli-Lubanski pseudo-vector 
\beq
L_\beta \equiv\eta_{\alpha\beta\mu\nu}U^\alpha (z_1^\mu(\tau_1) P_1^\nu+ z_2^\mu(\tau_2) P_2^\nu)
\eeq
is independent of $\tau_1$ and $\tau_2$, and
has, as only nonzero component, $L_z=  b\,P  = P_\phi= G m_1 m_2 j$ both in the ${\mathcal L}_1$ coordinate system,
and in the c.m. one.

\end{document}